\documentclass[%
 aip,
 amsmath,amssymb,
preprint,%
author-year,
]{revtex4-1}

\usepackage{graphicx}
\usepackage{dcolumn}
\usepackage{bm}

\usepackage[utf8]{inputenc}
\usepackage[T1]{fontenc}
\usepackage{mathptmx}
\usepackage{etoolbox}

\makeatletter
\def\@email#1#2{%
 \endgroup
 \patchcmd{\titleblock@produce}
  {\frontmatter@RRAPformat}
  {\frontmatter@RRAPformat{\produce@RRAP{*#1\href{mailto:#2}{#2}}}\frontmatter@RRAPformat}
  {}{}
}%
\makeatother
\begin{document}

\preprint{AIP/123-QED}

\title{The Parallel-Transported (Quasi)-Diabatic Basis}
\author{Robert Littlejohn}
 \email{robert@wigner.berkeley.edu}
 \affiliation{Department of Physics, University of California, Berkeley, California 94720 USA}
 
\author{Jonathan Rawlinson}%
 \email{jonathan.rawlinson@manchester.ac.uk}
\affiliation{School of Mathematics, University of Manchester, Manchester UK}

\author{Joseph Subotnik}
 \email{subotnik@sas.upenn.edu}
\affiliation{Department of Chemistry, University of Pennsylvania, 
Philadelphia, PA, USA}

\date{\today}

\begin{abstract}
  This article concerns the use of parallel transport to create a
  diabatic basis.  The advantages of the parallel-transported basis
  include the facility with which Taylor series expansions can be
  carried out in the neighborhood of a point or a manifold such as a
  seam (the locus of degeneracies of the electronic Hamiltonian), and
  the close relationship between the derivative couplings and the
  curvature in this basis.  These are important for analytic
  treatments of the nuclear Schr\"odinger equation in a neighborhood
  of degeneracies.  The parallel-transported basis bears a close
  relationship to the singular-value basis; in this article both are expanded in power
  series about a reference point and they are shown to agree through
  second order but not beyond.  Taylor series expansions are effected
  through the projection operator, whose expansion does not involve
  energy denominators or any type of singularity, and in terms of
  which both the singular-value basis and the parallel-transported
  basis can be expressed.  The parallel-transported basis is a version
  of Poincar\'e gauge, well known in electromagnetism, which provides
  a relationship between the derivative couplings and the curvature
  and which, along with a formula due to Mead, affords an efficient
  method for calculating Taylor series of the basis states and the
  derivative couplings.  The case in which fine structure effects are
  included in the electronic Hamiltonian is covered.
\end{abstract}

\maketitle

\newcommand{\AS}{{\mathcal{A}}}
\newcommand{\avec}{{\mathbf{a}}}
\newcommand{\Avec}{{\mathbf{A}}}
\newcommand{\bra}[1]{\langle#1\vert}
\newcommand{\braket}[2]{\langle#1\vert#2\rangle}
\newcommand{\Bregion}{{\mathcal{B}}}
\newcommand{\Bregionbar}{{\overline{\mathcal{B}}}}
\newcommand{\BS}{{\mathcal{B}}}
\newcommand{\Bvec}{{\mathbf{B}}}
\newcommand{\bvec}{{\mathbf{b}}}
\newcommand{\CaseIa}{{\hbox{\rm Case~Ia}}}
\newcommand{\CaseIb}{{\hbox{\rm Case~Ib}}}
\newcommand{\CasesIab}{{\hbox{\rm Cases~Iab}}}
\newcommand{\CaseII}{{\hbox{\rm Case~II}}}
\newcommand{\codim}{{\mathop{\textrm{codim}}}}
\newcommand{\Complexes}{\mathbb{C}}
\newcommand{\Cbar}{{\bar C}}
\newcommand{\CS}{{\mathcal{C}}}
\newcommand{\Ctilde}{{\tilde C}}
\newcommand{\HS}{{\mathcal{H}}}
\newcommand{\Integers}{\mathbb{Z}}
\newcommand{\iquat}{{\mathbf{i}}}
\newcommand{\jquat}{{\mathbf{j}}}
\newcommand{\Jvec}{{\mathbf{J}}}
\newcommand{\ket}[1]{\vert#1\rangle}
\newcommand{\ketbra}[2]{\vert#1\rangle\langle#2\vert}
\newcommand{\kquat}{{\mathbf{k}}}
\newcommand{\Ktilde}{{\tilde K}}
\newcommand{\Levels}{{A}}
\newcommand{\Lvec}{{\mathbf{L}}}
\newcommand{\matrixelement}[3]{\langle#1\vert#2\vert#3\rangle}
\newcommand{\MOI}{{\mathsf{M}}}
\newcommand{\Ne}{{N_e}}
\newcommand{\Nn}{{N_n}}
\newcommand{\nvechat}{{\hat{\mathbf{n}}}}
\newcommand{\omegavec}{\bm{\omega}}
\newcommand{\Pbar}{{\bar{P}}}
\newcommand{\Proj}{{\mathcal{P}}}
\newcommand{\Pvec}{{\mathbf{P}}}
\newcommand{\pvec}{{\mathbf{p}}}
\newcommand{\Quaternions}{\mathbb{H}}
\newcommand{\Reals}{\mathbb{R}}
\newcommand{\Region}{{\mathcal{R}}}
\newcommand{\Regionbar}{{\overline{\mathcal{R}}}}
\newcommand{\rvec}{{\mathbf{r}}}
\newcommand{\Rvec}{{\mathbf{R}}}
\newcommand{\scalarprod}[2]{\langle #1,#2\rangle}
\newcommand{\sigmavec}{\bm{\sigma}}
\newcommand{\Sspace}{{\cal{S}}}
\newcommand{\Svec}{{\mathbf{S}}}
\newcommand{\thetavec}{\bm{\theta}}
\newcommand{\tr}{{\mathop{\textrm{tr}}}}
\newcommand{\Vvec}{{\mathbf{V}}}
\newcommand{\xbar}{{\bar x}}
\newcommand{\xtilde}{{\tilde x}}
\newcommand{\xvec}{{\mathbf{x}}}
\newcommand{\Xvec}{{\mathbf{X}}}
\newcommand{\Xvectilde}{\tilde{\mathbf{X}}}
\newcommand{\ytilde}{{\tilde y}}
\newcommand{\yvechat}{{\hat{\mathbf{y}}}}
\newcommand{\Yvec}{{\mathbf{Y}}}
\newcommand{\Yvectilde}{\tilde{\mathbf{Y}}}
\newcommand{\ztilde}{{\tilde z}}
\newcommand{\Zvectilde}{\tilde{\mathbf{Z}}}

\section{Introduction}
\label{intro}

This article concerns bases of electronic wave functions in
polyatomic molecules that span a chosen subspace of strongly coupled
states.  We are especially interested in the parallel-transported
basis and the closely related singular-value basis.  The context of
this work is Born-Oppenheimer theory, applied to multiple, strongly
coupled electronic states, a subject that is treated and reviewed in
many places (\cite{Cederbaum04, Farajietal12, Koppel04a,
MatsunagaYarkony98, Yarkony96, Yarkony04b, Yarkony01, Yarkony12,
ZhuYarkony16, Kendricketal02, Mead88, Mead92, MeadTruhlar79,
MeadTruhlar82, RichingsWorth15, RichingsHabershon20}).  The bases we refer to are really fields of bases,
that is, defined over some region of the nuclear configuration space.

In this article we refer to any basis that is smooth over the region
in question as ``diabatic,'' without any assumption that the
derivative couplings vanish and without the common prefix ``quasi.''
We do this because as a practical matter the derivative couplings in
polyatomic molecules never vanish; if we wish to refer to the (rare) cases
in which they do, we will refer to a ``strictly diabatic'' basis.  In
addition, the designation ``quasi'' implies that a quasi-diabatic
basis is close to a strictly diabatic basis, that is, that the
derivative couplings are small.  There is much wishful thinking in the
literature that the derivative couplings can be made small enough to
be ignored, but without quantitative justification.  A careful
analysis by \textcite{Kendricketal02}, however, shows that the
derivative couplings are generically of order unity over a range of
the order of one atomic unit, unless forced to be even larger by small
energy denominators.  In addition numerical evidence
(\cite{ChoiVanicek21}) supports the conclusion that these couplings must
be taken into account for accurate results.

There is some subtlety in the question of the magnitude of the
derivative couplings, however, because these couplings are components
of a vector potential of a gauge theory (\cite{Pacheretal89,
Pacheretal93, Wittig12, KendrickMead95, Mead80b, Mead87, Mead92,
Bohmetal91, Bohmetal92, Bohmetal92a, Kendrick04, Berry84}), in which
the gauge transformations are changes of frame, specified by (ideally)
smooth fields of unitary matrices.  We call this the
Mead-Truhlar-Berry vector potential or connection
(\cite{MeadTruhlar79,Berry84}).  In this article, the derivative
couplings and the Mead-Truhlar-Berry connection or vector potential
(the only vector potential we consider) are almost the same thing (the
small differences are explained in Sec.~\ref{DerivativeCandConnection}).

Thus, diabatic bases are not unique, and different choices have
different advantages and disadvantages. Several choices of diabatic
basis have been reviewed by \textcite{Pacheretal93}, including the
singular-value basis (\cite{Pacheretal88}) and the Lorentz-gauge basis
(\cite{Pacheretal89}), which minimizes the mean square of the connection
over a region of nuclear configuration space.
  
In addition there is the basis of \textcite{WernerMeyer81}, which
diagonalizes the dipole operator in a single dimension, and of
\textcite{CaveNewton96,CaveNewton97} and
\textcite{Subotniketal08} who generalized this
diagonalization to multidimensional problems.  Historically, within
the chemistry literature, the emphasis has been on finding a good set
of diabatic states that serve as initial and final states for electron
transfer problems, e.g., \textcite{Subotniketal09}, so
that one can apply Marcus theory; that being said, our treatment here
is more general dynamically (rather than serving as a platform for a
spin-boson model Hamiltonian (\cite{XuSchulten94,RossoDupuis06})). In
such a case, there is no need to have an element of electronic
locality inherent within a smooth diabatic basis. 

The advantages of the parallel-transported basis include the facility
with which it can be expanded in Taylor series about given points or
even given manifolds, such as seams or degeneracy manifolds, and
the close relationship it bears to the curvature.  These features will
be explained in detail in this article, but the following is some of
the basic ideas.

It turns out that by means of a gauge transformation it is possible to
make the vector potential vanish at any given point.  Indeed, both the
parallel-transported basis and the singular-value basis do this, as
will be explained below.  In fact, in some circumstances (triatomics
in the electrostatic model) there exists a parallel-transported frame
that causes the vector potential to vanish over the entire degeneracy
manifold or seam, which is a one-dimensional curve.  In the
neighborhood of such a point or such a manifold in such a basis, the
derivative couplings are indeed small.  Then the question is how
rapidly the couplings grow as we move away from the chosen point or
manifold.

This question draws attention to the curvature, a tensor that is a
function of the derivative couplings and their derivatives.  Here the
important fact is that the curvature tensor, unlike the vector
potential, is a {\em tensor}, so that it transforms under a gauge
transformation by simple conjugation by the unitary matrix specifying
the transformation, see (\ref{Gxfm}).  Thus, the magnitude of the
curvature tensor, whose square is the sum of squares of its
components, is invariant under a gauge transformation.  In a sense the
curvature represents the part of the connection that is
gauge-invariant, while the part that is not is what is changed by a
gauge transformation.  We will not attempt to make this notion precise
but merely use it to motivate the idea that to minimize the connection
we should find a gauge that expresses the connection in terms of the
curvature.   As we will show, the parallel-transported basis does
this. 

Our interest in the parallel-transported basis is part of a larger
project in which we aim to find analytical treatments of the nuclear
Schr\"odinger equation in the neighborhood of conical intersections.
Part of that work involves normal forms for Landau-Zener transitions
in many dimensions; these are illustrated by
\textcite{LittlejohnFlynn92}, in which a 2-dimensional Landau-Zener
problem is treated.  A conical intersection has an effective range in
the nuclear configuration space, outside of which the potential energy
surfaces can be decoupled by adiabatic means
(\cite{WeigertLittlejohn93,Panatietal02,Teufel03}).  That range depends
on the nuclear momentum and gets larger as that momentum grows, but
even for fairly large nuclear momenta by ordinary standards, that is,
momenta such that the nuclear kinetic energy is of the order of one
atomic unit, the range of a conical intersection is still quite small
in atomic units.  Thus, Taylor series expansions are a viable approach
for covering the strongly coupled region.

Expansions in Taylor series about a conical intersection have a long
history, including the works by
\textcite{Mead83,ThompsonMead85,Yarkony97a} on derivative couplings
and their singularities in the neighborhood of a conical intersection
and by \textcite{Yarkony00} on the bifurcations of degeneracy
manifolds.  This work focuses on the projection operator and its
expansion, an attractive intermediate goal since the projection
operator is smooth and diabatic bases such as the singular-value and
parallel-transported basis can be expressed in terms of it.  The usual
approaches work with the adiabatic basis, which leads to small or
vanishing energy denominators when one moves tangent or nearly tangent
to the degeneracy manifold or seam.  In this article we avoid the
adiabatic basis as much as possible, mainly using the projection
operator as a substitute; this gives us expansions that are valid even
when one moves tangent to the degeneracy manifold. 

In Sec.~\ref{singularvaluebasis} we discuss aspects of the
singular-value diabatic basis, which is due to
\textcite{Pacheretal88,Pacheretal93}.  We begin with a variational
principle for the singular-value diabatic basis that differs somewhat
from the one used in the original work on the subject, but which shows
a closer connection with the variational foundations of the
parallel-transported basis, which is our main topic.  We emphasize the
issue of the smoothness of the solution; although the singular-value
diabatic basis is smooth, some of the the objects used in its
derivation and construction are not.  We express the singular-value
basis in terms of the projection operator, an important step since the
latter is smooth.  In addition, the projection operator can be
expanded about a reference point, possibly on a conical intersection,
something that we carry out in Sec.~\ref{expansionprojection}.  We
then use that to expand the singular-value basis to third order about
the reference point.

In Sec.~\ref{diabaticparallel} we develop the theory of parallel
transport and use it to construct the parallel-transported basis and
its expansion about a reference point, which we show agrees with the
singular-value basis through second order but differs at third order.
We then express the Hamiltonian, connection and curvature in the
parallel-transported bases as power series expansions about a
reference point.  We point out that the parallel-transported frame is
a version of Poincar\'e gauge, which in electromagnetism is a gauge
that is transverse in real space.  This allows us to express the
connection in terms of an integral involving the curvature, which
provides an efficient method of calculating the connection in the
parallel-transported frame.  Finally we point out that constructions
of diabatic bases are only needed in $3N-3$ directions of nuclear
configuration space, because in the remaining 3 directions the nuclear
configuration changes by a pure rotation, and in those directions the
basis should transform by a rotation operator.   Thus the construction
of diabatic bases takes place on a $(3N-6)$-dimensional ``section'' of
the nuclear configuration space.

In Sec.~\ref{degeneracymanifolds} we discuss degeneracy manifolds,
usually called seams, and the construction of diabatic bases in the
neighborhood of them.  For example, in triatomic molecules in the
electrostatic model for the electronic Hamiltonian, for which the
degeneracy manifold is a 1-dimensional curve in the 3-dimensional
section, we construct a diabatic basis in a tubular region surrounding
the curve.  Again the connection is expressed in terms of the
curvature, but the components of the connection along the degeneracy
manifold and transverse to it have different expansions.  We also
point out that in the case of triatomics with the electrostatic model,
it is possible to choose a gauge such that the derivative couplings
(in all of their components) vanish on the degeneracy manifold.
Finally, in Sec.~\ref{conclusions} we present some conclusions. 

In this article we denote the adiabatic basis by $\ket{ax;k}$, the
singular-value diabatic basis by $\ket{dx;k}$, the basis that is parallel
transported away from a reference point by $\ket{px;k}$ and the basis
that is parallel transported away from the degeneracy manifold by
$\ket{Dx;k}$, where $x$ is a point of nuclear configuration space and
$k$ is a quantum number or sequencing number for the basis state.

\section{The Singular-Value Diabatic Basis}
\label{singularvaluebasis}

In this section we cover some aspects of the singular-value diabatic
basis, which is due to \textcite{Pacheretal88,Pacheretal93}.  This
basis is easy to compute at a point $x$ of configuration space, and it
does not require integration along some path starting at another point
$x_0$.  For these reasons it is popular in numerical work.  It is also
conceptually simple and geometrically compelling.  In this article we
present some new results on the singular-value basis, including a
discussion of its smoothness properties, its relation to the
projection operator, and Taylor series expansions about a reference
point.

We begin
by showing that the singular-value basis satisfies a certain variational principle.
This differs only slightly from the original variational principle used by
\textcite{Pacheretal93}, but we present it anyway because it reveals a
close connection with parallel transport which is the main topic of
this paper.  Both variational principles lead to the same
(singular-value) diabatic basis.  Other variational principles that
have been considered, for example, the one by
\textcite{Cimiragliaetal85}, are not equivalent and lead to
different diabatic bases.

We also emphasize the issue of the smoothness of the singular-value
diabatic basis, something that takes some effort since the adiabatic
basis with which the construction begins is not smooth.  Finally we
show that the singular-value diabatic basis can be expressed in terms
of the projection operator onto the coupled subspace of the electronic
Hilbert space.  Since the projection operator is smooth, this shows
that the singular-value diabatic basis is, too, and it also provides a
method for expanding the singular-value basis in a Taylor series about
a reference point.

\subsection{Terminology and Notation}  

We let $x$ represent a point of the nuclear configuration space in the
center-of-mass frame, so that $x$ stands for the collection of $N-1$
Jacobi vectors (\cite{AquilantiCavalli86,SmirnovShitikova77}) or their
$3(N-1)$ components, where $N\ge3$ is the number of nuclei.  We let
$H(x)$ be the electronic Hamiltonian, which initially we assume is
taken in the electrostatic model, although later we make some comments
about the changes necessary when fine structure effects are included.
We label the energy levels of $H(x)$ by a sequencing number $k$, and
define a ``coupled subspace'' by a set of adjacent energy levels,
\begin{equation}
  \Levels=\{k_0,k_0+1,\ldots,k_0+N_l-1\},
  \label{Idef}
  \end{equation}
where $N_l$ is the number of coupled levels.  Usually in practice
$k_0$ is the ground state and $N_l$ is small.  We consider a region
of nuclear configuration space in which energy level $k_0$ is not
degenerate with level $k_0-1$ and level $k_0+N_l-1$ is not degenerate
with level $k_0+N_l$, that is, a region in which there are no
degeneracies that cross the boundaries of $\Levels$.  Degeneracies within
$\Levels$ (``internal degeneracies'') are allowed, however (and these
require $N_l\ge2$).  In practice, if the region we are interested in
does have degeneracies that cross the boundaries of $\Levels$ (caused
by ``intruder states'' (\cite{VenghausEisfeld16})), we can either
excise the locations where those degeneracies occur from the region in
question, or we can expand the definition of $\Levels$ to make the
problematic degeneracies internal. If we choose the first option then
the region after the excision will often fail to be simply connected,
which has topological implications for the existence of smooth fields
of frames.

For $k\in \Levels$ we denote the energy eigenstates by $\ket{ax;k}$, so that
\begin{equation}
  H(x)\ket{ax;k}=\epsilon_k(x)\ket{ax;k},\qquad k\in \Levels,
  \label{energyeigenkets}
  \end{equation}
where, as indicated, the energy eigenvalues are $\epsilon_k(x)$.  The
$a$ in the notation $\ket{ax;k}$ indicates the adiabatic basis, and
the $x$-dependence of the eigenstate is separated by a semicolon from
the quantum number $k$.

We let $\Sspace(x)$ be the subspace of the electronic Hilbert space
spanned by the set of vectors $\ket{ax;k}$ for $k\in \Levels$, what we will
call the ``coupled subspace.''  The projection operator onto this
subspace is
\begin{equation}
  P(x)=\sum_{k\in \Levels}\ketbra{ax;k}{ax;k}.
  \label{Pdef}
  \end{equation}
We denote the subspace of the Hilbert space orthogonal to $\Sspace(x)$
by $\Sspace^\perp(x)$, and denote the projection operator onto
$\Sspace^\perp(x)$ by
\begin{equation}
  Q(x)=1-P(x).
  \label{Qdef}
  \end{equation}

For $k\notin \Levels$ we write $\ket{ax;k}$ for a basis in
$\Sspace^\perp(x)$ that is discrete and orthonormal but otherwise
arbitrary. We will call the set of states $\{\ket{ax;k}\}$ for all $k$
the ``adiabatic basis'', even though the basis states for $k\notin \Levels$
are not energy eigenstates.  This is not exactly standard terminology
but in fact normal usage is often restricted to the states $k\in \Levels$.
In addition there are reasons not to deal with energy eigenstates when
$k\notin \Levels$, which present a number of difficulties, both
computational and theoretical.  Computationally the high lying energy
eigenstates are seldom accessible, and theoretically they present
problems because their phase and frame conventions have singularities
on surfaces that proliferate at $k$ is increased.  In addition there
is the problem of the continuum and the transition thereto, which
changes as $x$ changes.  Actually one never needs to refer to energy
eigenstates when $k\notin \Levels$ and we shall not do so.  We must
remember, however, that in our ``adiabatic basis'' the vectors
$\ket{ax;k}$ are energy eigenstates only when $k\in \Levels$.

\subsection{Phase Conventions and Smoothness}

Energy eigenstates are defined only to within a phase convention (when
nondegenerate) or a frame convention inside the degenerate eigenspace
(when degenerate), and this must be kept in mind when using the
notation $\ket{ax;k}$ for $k\in \Levels$.  Although it is always possible to
assign those conventions at any given point $x$ of the region in
question, in general this assignment is not smooth on certain
submanifolds of the region as $x$ is varied.  Notably this occurs when
$x$ lies on an internal degeneracy (usually a conical intersection).
There may be additional singularities when $x$ is not on an internal
degeneracy; these occur on surfaces that can be moved about by a gauge
transformation like a branch cut in the complex plane or the strings
of monopoles but that cannot be eliminated.  These singularities (of
both types) are the main drawback of the adiabatic basis.

In this article when we say that something is smooth we mean that it
is a continuous function of $x$ and that it has as many continuous
derivatives as needed for any applications we might make.  A function
is continuous at a point $x_0$ if it is defined at $x_0$ and its limit
as $x\to x_0$ is independent of the direction of approach and equals
the value of the function at $x_0$.  It is precisely in this sense
that the adiabatic basis is not continuous at an internal degeneracy
(the limit exists as $x\to x_0$ but it depends on the direction of
approach), and the lack of continuity implies that the derivatives of
the adiabatic basis diverge at such points. 

In fact the singularities of the adiabatic basis occur only on a
subset of measure zero, so elsewhere, on a subset of full measure, it
is possible to talk about the derivatives of the adiabatic basis
vectors, as is commonly done in the literature.  Such derivatives
occur in the derivative couplings and other quantities of interest.
Such quantities diverge as a degeneracy is approached, and a great
deal of literature is devoted to analyzing and eliminating these
divergences.  In this article we will simply refrain from ever
assuming that the vectors $\ket{ax;k}$ are smooth functions of $x$,
and, in particular, we never talk about derivatives of these vectors
with respect to $x$.  This applies both for $k\in \Levels$ and $k\notin \Levels$;
at any given $x$ we can choose $\ket{ax;k}$ for $k\notin \Levels$ as a basis
that spans $\Sspace^\perp(x)$, but we will not assume that that
assignment can be made in a smooth manner as $x$ is varied.

The singularities of the adiabatic basis vectors $\ket{ax;k}$ for
$k\in \Levels$ are always due the impossibility of making phase and/or frame
conventions in a smooth manner as $x$ is varied, but the projection
operator $P(x)$ is independent of those conventions and is always
smooth.  (The projection operator is not defined where degeneracies cross the boundaries of the set $A$, but we are excluding such points from the region under consideration.)  This means that the derivatives of $P(x)$ exist and that
$P(x)$ can be expanded in a Taylor series about any point $x_0$, in
which all the derivatives are well defined.  We will present such an
expansion below.  Since $P(x)$ is smooth, $Q(x)=1-P(x)$ is too.

\subsection{Diabatic Bases are Smooth}

Now given a configuration $x_0$ we define a diabatic basis as one that
is smooth (free of singularities) in a neighborhood of $x_0$. In the
literature one is usually interested in the case that $x_0$ lies on a
degeneracy manifold (usually a seam or surface of conical
intersection) but there are applications for which this is not so and
in the following we shall make no assumptions about $x_0$.  A diabatic
basis so defined is not unique, since any smooth field of unitary
transformations on $\Sspace(x)$ and another on $\Sspace^\perp(x)$ will
map one smooth basis into another.  As is well known
(\cite{Pacheretal89,Bohmetal92,Mead92,Pacheretal93}), such changes of
frame are gauge transformations of the theory.  Thus additional
criteria must be imposed for the selection of a unique diabatic basis.

To characterize a diabatic basis as one that is smooth does not
completely capture the practical requirements of such a basis.  One
could  take a smooth diabatic basis and subject it to a gauge
transformation (a change of frame) based on a unitary transformation
that was a smooth function of $x$ in the mathematical sense but
rapidly varying on the atomic length scale.   The resulting basis
would still be smooth in a mathematical sense but it would have large
derivative couplings and not be useful.  So a diabatic basis must not
only be smooth it must be slowly varying, that is, it should have a
variation of  order unity on the atomic length scale.  One can
formalize this in terms of the dependence on the Born-Oppenheimer
ordering parameter $\kappa=(m/M)^{1/4}$; the diabatic basis should not
depend on $\kappa$.   In  the following, such understandings about the
word ``smooth'' will be assumed.

For reference we will need a basis in the electronic Hilbert space
that is independent of $x$.  For convenience we choose this basis to
be $\ket{ax_0;k}$, which is the adiabatic basis at $x_0$.  In the
literature when this basis is used to study dynamics at points $x\ne
x_0$ it is sometimes called the ``crude adiabatic basis.''  

\subsection{A Variational Criterion}

We now present a variational criterion for a diabatic basis, which
leads to the singular-value basis.  Initially we focus on the part
$k\in \Levels$ of the basis that spans the coupled subspace $\Sspace(x)$.
We define the diabatic basis at $x$, denoted $\ket{dx;k}$ with a
designation $d$ for ``diabatic,'' as the frame at $x$ that is closest
to the adiabatic frame at $x_0$, in the sense that it minimizes the
quantity
\begin{equation}
  \sum_{k\in \Levels} \Big|\ket{dx;k} - \ket{ax_0;k}\Big|^2=
  \sum_{k\in \Levels} (2 - \braket{ax_0;k}{dx;k} - \braket{dx;k}{ax_0;k}).
  \label{framedistsq}
  \end{equation}
Both the diabatic basis $\ket{dx;k}$ and the adiabatic one
$\ket{ax;k}$ for $k\in \Levels$ span $\Sspace(x)$, so we must have
\begin{equation}
  \ket{dx;k}=\sum_{l\in \Levels} \ket{ax;l}\, U_{lk}(x),
  \label{diabdef}
  \end{equation}
where $U_{lk}(x)$ is an $N_l\times N_l$ unitary matrix.   Thus the
quantity (\ref{framedistsq}) can be written,
\begin{equation}
  \sum_{k\in \Levels} \left[2 - \sum_{l\in \Levels}( S_{kl} \, U_{lk}
    + S^*_{kl} U^*_{lk})\right],
  \label{framedist}
  \end{equation}
where
\begin{equation}
  S_{kl}=\braket{ax_0;k}{ax;l}.
  \label{Sdef}
  \end{equation}
The notation $S$ for this matrix is the same as in
\textcite{Pacheretal93}.

With no restrictions on $k$ and $l$, (\ref{Sdef}) defines $S$ as an
infinite-dimensional, unitary matrix, but for use in the variational
principle we only need the $N_l\times N_l$ block of this matrix,
$S_{kl}$ for $k,l\in \Levels$.  We find it convenient to partition infinite
dimensional matrices like this according to what we will call the $a$
and $b$ subsets of indices, which are those for which $k\in \Levels$ and
$k\notin \Levels$, respectively.  For example, we partition the infinite
dimensional matrix $S$ as follows:
\begin{equation}
  S=\left(\begin{array}{cc}
    S^{aa} & S^{ab} \\
    S^{ba} & S^{bb}
    \end{array}\right),
  \label{Spartition}
  \end{equation}
so that $S^{aa}$ is an $N_l \times N_l$ matrix, etc.  In terms of this
notation, the quantity (\ref{framedistsq}) or (\ref{framedist}) can be
written
\begin{equation}
  2N_l - [\tr(S^{aa}U) + \hbox{\rm c.c.}].
  \end{equation}

We assume that the adiabatic basis is given as a function of $x$ so
that the matrix $S^{aa}$ is known (but it is not assumed to be a smooth
function of $x$).  We wish to minimize the quantity (\ref{framedist})
with respect to the choice of the unitary matrix $U_{kl}$ (where
$k,l\in \Levels$).  We enforce the unitarity of $U$ by adding the term,
\begin{equation}
  \tr[\Lambda(UU^\dagger-I)]=\sum_{k,l\in \Levels}
  \Lambda_{lk} \left(\sum_{m\in \Levels} U_{km}U^*_{lm}-\delta_{kl}\right),
  \label{LMterm}
  \end{equation}
where $\Lambda_{kl}$ is a Hermitian matrix of Lagrange multipliers.
Now varying with respect to either $U$ or $U^\dagger$ we find
\begin{equation}
  S^{aa}=U^\dagger\Lambda,
  \label{Ssoln}
  \end{equation}
in which $S^{aa}$ is given and unitary $U$ and Hermitian $\Lambda$ are
to be determined.  The theory of the polar decomposition tells us that
if $S^{aa}$ is nonsingular then (\ref{Ssoln}) has a unique solution
for a unitary matrix $U$ and a positive definite, Hermitian matrix
$\Lambda$.  We will show below that $S^{aa}$ is, in fact, nonsingular
in a neighborhood of $x_0$ and that $\Lambda$ must be positive
definite for a meaningful solution; thus we obtain unique results for
$U$ and $\Lambda$ over the neighborhood in question.  That is, since
$(S^{aa})^\dagger S^{aa}=\Lambda^2$, $\Lambda=[(S^{aa})^\dagger
S^{aa}]^{1/2}$, where here and below by the square root of a positive
definite, Hermitian matrix we mean the unique positive definite,
Hermitian square root; and given $\Lambda$, $U=\Lambda(S^{aa})^{-1}$.
Then (\ref{diabdef}) gives us the diabatic basis at $x$ in terms of a
finite linear combination of adiabatic basis vectors at $x$.  This is
identical to the singular-value diabatic basis of
\textcite{Pacheretal88}.

\subsection{Smoothness of the Solution}

This construction does not, however, make it evident that the
resulting diabatic basis is a smooth function of $x$.  In fact, it is
smooth, but neither $S^{aa}$, $U$ or $\Lambda$ are.  Since
singularities are due to the phase and frame conventions for the
adiabatic basis states $\ket{ax;k}$ (frame conventions applying inside
a degenerate subspace), let us change these conventions for $k\in \Levels$
by means of an $N_l\times N_l$ unitary matrix $V$,
\begin{equation}
  \ket{a'x;k} = \sum_{l\in \Levels} \ket{ax;l} V_{lk},
  \label{achange}
  \end{equation}
and study how various quantities transform under such a change.  In
(\ref{achange}) $V$ is allowed to be a nonsmooth function of $x$.
Combining (\ref{achange}) with (\ref{diabdef}), the definition of $U$,
and with (\ref{Sdef}), the definition of $S^{aa}$, we find
\begin{equation}
  S^{aa\prime}=S^{aa}V, \qquad U'=V^\dagger U, \qquad 
  \Lambda'=V^\dagger\Lambda V.
  \end{equation}
Thus neither $S^{aa}$, $U$ nor $\Lambda$ is smooth, in general.

On the other hand, the products $S^{aa}U=U^\dagger \Lambda U$ and
$S^{aa}(S^{aa})^\dagger=U\Lambda^2U$ are invariant under the change
(\ref{achange}) and are therefore smooth (for example,
$S^{aa\,\prime}U'=S^{aa}U$), as is the quantity $|\det S^{aa}|$.  As
for the latter, we note that at $x=x_0$, $S^{aa}=I$ so $|\det
S^{aa}|=1$.  Therefore by continuity $\det S^{aa}$ is bounded away
from 0 in some finite neighborhood of $x_0$ (and note that we can use
continuity arguments only on smooth functions), and therefore $S^{aa}$
is nonsingular in this neighborhood.

Let us define
\begin{equation}
  \Sigma = S^{aa}U = U^\dagger \Lambda U=\Sigma^\dagger,
  \label{Sigmadef}
  \end{equation}
which, as noted, is a smooth function of $x$.  To get its value at
$x_0$ we require of our diabatic basis that
$\ket{dx_0;k}=\ket{ax_0;k}$, that is, it agree with the adiabatic
basis at $x_0$.   This means that $U=I$ at $x=x_0$, and thus also
$\Lambda=\Sigma=I$ at $x_0$.   But since $\Sigma$ is a smooth function
of $x$, it must be positive definite in a neighborhood of $x_0$.  This
means that $\Lambda$, which is not smooth, is also positive definite
in the same neighborhood.

These considerations suggest that rather than factoring $S^{aa}$ via
the polar decomposition as in (\ref{Ssoln}), we factor it the other
way using $S^{aa} = \Sigma U^\dagger$ to obtain $\Sigma$ and $U$.
That is, we compute $\Sigma =[S^{aa}(S^{aa})^\dagger]^{1/2}$, and then
compute $U=(S^{aa})^{-1}\Sigma$.   The advantage of this approach is that $\Sigma$, a smooth
function of $x$, takes  the place of the nonsmooth $\Lambda$, which is
never needed.

\subsection{Relation to the Projection Operator}

We still have not shown that the diabatic basis $\ket{dx;k}$ is a
smooth function of $x$, but if it is we guess that it should be
possible to express it purely in terms of the projection operator
$P(x)$, which is known to be smooth.  This turns out to be the case,
as we now show.

For $k\in \Levels$ the procedure above gives the diabatic basis $\ket{dx;k}$
as a finite linear combination of the adiabatic basis $\ket{ax;k}$,
also for $k\in \Levels$.  For some purposes, however, it is convenient to
have the diabatic basis expressed as a linear combination of the
reference basis $\ket{ax_0;k}$, in spite of the fact that this
involves an infinite number of terms.  We obtain a useful form of this
expansion by writing, for $k\in \Levels$,
\begin{eqnarray}
  \ket{dx;k} &=& P(x)\ket{dx;k} = \sum_{{\rm all}\;l}
  \ketbra{ax_0;l}{ax_0;l} \sum_{m\in \Levels} 
  \ket{ax;m}\braket{ax;m}{dx;k}\nonumber\\
  &=&\sum_{l,m\in \Levels} \ket{ax_0;l} S^{aa}_{lm} U_{mk}+
  \sum_{l\notin \Levels,m\in \Levels} \ket{ax_0;l}S^{ba}_{lm}U_{mk}
  \nonumber\\
  &=& \sum_{l\in \Levels}\ket{ax_0;l}(S^{aa}U)_{lk} +
  \sum_{l\notin \Levels}\ket{ax_0;l} (S^{ab}U)_{lk}.
  \label{dxkxpn}
  \end{eqnarray}
Evidently, for this expansion we need the matrices $S^{aa}U=\Sigma$
and $S^{ba}U$.

We define the matrix representation of the projection operator $P(x)$
in the reference basis,
\begin{equation}
  P_{kl}(x) = \matrixelement{ax_0;k}{P(x)}{ax_0;l}=
  \sum_{m\in \Levels} \braket{ax_0;k}{dx;m}\braket{dx;m}{ax_0;l},
  \label{Pkldef}
  \end{equation}
for all $k$, $l$.  Now substituting (\ref{dxkxpn}) into
(\ref{Pkldef}), we obtain
\begin{eqnarray}
  P^{aa} &=&(S^{aa}U)(S^{aa}U)^\dagger= S^{aa}(S^{aa})^\dagger = \Sigma^2,
  \nonumber\\
  P^{ba} &=& (S^{ba}U)(S^{aa}U)^\dagger= S^{ba}(S^{aa})^\dagger,
  \label{P1stcol}
  \end{eqnarray}
where we partition $P(x)$ as in (\ref{Spartition}).  Thus the matrix
$S^{aa}U$ that appears in (\ref{dxkxpn}) is the positive definite
matrix $\Sigma=(P^{aa})^{1/2}$.  As for the matrix $S^{ba}U$, we have
\begin{equation}
  P^{ba} = S^{ba}U (S^{aa}U)^\dagger=S^{ba}U\Sigma^\dagger = 
  S^{ba}U\Sigma,
  \end{equation}
or,
\begin{equation}
  S^{ba}U = P^{ba}(P^{aa})^{-1/2}.
  \label{Tbasoln}
  \end{equation}
Thus, both matrices needed in the expansion (\ref{dxkxpn}) can be
expressed in terms of the projection operator.

Let us write $T$ for the infinite-dimensional unitary matrix that
connects the diabatic basis $\ket{dx;k}$ with the reference basis
$\ket{ax_0;k}$, 
\begin{equation}
  \ket{dx;k} = \sum_{{\rm all}\;l}\ket{ax_0;l}T_{lk},
  \label{Tdef}
  \end{equation}
which holds for all $k$, $l$.  Again, this notation follows
\textcite{Pacheretal88,Pacheretal93}, and we note that their
variational principle minimizes the quantity
\begin{equation}
  \big|T-I\big|^2.
  \end{equation}
So far we have worked on the case $k\in \Levels$ and found $T^{aa}=S^{aa}U$
and $T^{ba}=S^{ba}U$ in terms of the projection operator.  We can find
the other blocks, $T^{ab}$ and $T^{bb}$, by repeating the same
procedure we have just been through, but now taking $k\notin \Levels$ and
swapping subspaces $a\leftrightarrow b$ and projectors
$P\leftrightarrow Q$.  This gives altogether,
\begin{eqnarray}
  T &=& \left(\begin{array}{cc}
      (P^{aa})^{1/2} & Q^{ab}(Q^{bb})^{-1/2} \\
      P^{ba}(P^{aa})^{-1/2} & (Q^{bb})^{1/2}
      \end{array}\right)\nonumber\\
    &=&\left(\begin{array}{cc}
        P^{aa} & Q^{ab} \\
        P^{ba} & Q^{bb}
        \end{array}\right)
      \left(\begin{array}{cc}
          (P^{aa})^{-1/2} & 0 \\
          0 & (Q^{bb})^{-1/2}
          \end{array}\right),
	\label{TintermsofP}
  \end{eqnarray}
where we note that $Q^{ab}=-P^{ab}$ and $Q^{bb}=I^{bb}-P^{bb}$, so
that the entire result can be expressed in terms of $P$.  It can be
directly checked that $TT^\dagger = T^\dagger T = I$; for this one
must use the facts that $P=P^\dagger$ and $P^2=P$.

Now (\ref{Tdef}) and (\ref{TintermsofP}) show that the transformation
from the reference basis to the singular-value diabatic basis can be
expressed purely in terms of the projection operator.  This not only
shows that the diabatic basis is smooth in a neighborhood of $x_0$ but
it also allows us to expand the diabatic basis in a Taylor series
about $x_0$.

\section{Expansion of the Projection Operator}
\label{expansionprojection}

In the following we will use the symbol $P$ or $P(x)$ for both the
projection operator and its matrix in the reference basis, as in
(\ref{Pkldef}), with hopefully little danger of confusion.  We wish to
expand $P$ in a power series about the reference point $x_0$.  We
write $x=x_0+\xi$, where $\xi$ is a displacement vector in nuclear
configuration space.  We think of all the symbols, $x$, $x_0$ and
$\xi$, as standing for the $N-1$ Jacobi vectors or their $3(N-1)$
components, effectively using these as coordinates on the nuclear
configuration space in the center-of-mass system.   We will write
these coordinates explicitly as $x^\mu$, $x_0^\mu$, $\xi^\mu$, etc.,
where $\mu=1,\ldots,3N-3$.   We use the summation convention on such
coordinate indices $\mu$, $\nu$, etc. 

The expansion of the projection operator is a part of degenerate
or quasidegenerate perturbation theory, see for example
\textcite{Kato49,Bloch58,DesCloizeaux60,Klein74}, which discuss, among
other techniques, the use of Cauchy's theorem and the resolvent
operator to obtain expansions of the projection operator.  In the
following we summarize a straightforward approach, which is suitable
to the order to which we carry the expansion.  We simply note that the
usual goal of perturbation theory is to expand the energy eigenstates
(which in our case would be the adiabatic basis) in a power series,
something which however must cope with degeneracies, near degeneracies
and small or vanishing energy denominators, none of which is an issue
in the expansion of the projection operator.  These complications are
due to the singularity of the adiabatic basis at a degeneracy, which
we are avoiding by using the projection operator.

We expand the projection operator in a power series in $\xi$,
\begin{equation}
  P(x)=P_0 + P_1 + P_2 + \ldots,
  \label{Pexpand}
  \end{equation}
where
\begin{equation}
  P_0 = P(x_0), \;
  P_1 = \xi^\mu \frac{\partial P}{\partial x^\mu}(x_0), \;
  P_2 = \frac{1}{2} \xi^\mu \xi^\nu \frac{\partial^2 P}{\partial x^\mu
    \partial x^\nu}(x_0),
  \label{Pndefs}
  \end{equation}
etc., and similarly we expand the Hamiltonian,
\begin{equation}
  H(x) = H_0 + H_1 + H_2 + \ldots.
  \label{Hexpand}
  \end{equation}

We carry out the perturbation expansion by requiring
$P(x)^\dagger=P(x)^2=P(x)$ and $[P(x),H(x)]=0$.   At zeroth order,
that is, at $x=x_0$, we have
\begin{equation}
  P_0=\left(\begin{array}{cc}
      I^{aa} & 0 \\
      0 & 0
      \end{array}\right),
    \qquad
    H_0 = \left(\begin{array}{cc}
        H_0^{aa} & 0 \\
        0 & H_0^{bb}
        \end{array}\right),
      \label{PH0defs}
      \end{equation}
where matrices are partitioned as in (\ref{Spartition}).  Since the
reference basis is an energy eigenbasis at $x_0$ for $k\in \Levels$, we have
\begin{equation}
  H^{aa}_{0,kl} = \epsilon_{0k}\,\delta_{kl},
  \label{Haa0def}
  \end{equation}
where $\epsilon_{0k} = \epsilon_k(x_0)$ for $k\in \Levels$.
Equation~(\ref{Haa0def}) applies when $k,l\in \Levels$; in equations like
this we shall not indicate the ranges of indices when they are evident
from the superscripts (for example, $aa$ in this case).

At first order we must satisfy $P_0P_1+P_1P_0=P_1$, which gives
$P^{aa}_1=0$ and $P^{bb}_1$=0.  In addition, when we require
$[P_0,H_1] + [P_1,H_0]=0$, we find
\begin{equation}
  P^{ab}_{1,kl} = \sum_{m\notin \Levels} H^{ab}_{1,km}
  R(\epsilon_{0k})_{ml},
  \label{P1absoln}
  \end{equation}
where the resolvent $R(\epsilon)$ is a $bb$-type matrix defined by
\begin{equation}
  \sum_{m\notin \Levels} R(\epsilon)_{km} \big(\epsilon\,
  \delta_{ml} - H^{bb}_{0,ml}\big)
  =\delta_{kl},
  \label{Rdef}
  \end{equation}
that is,
\begin{equation}
  R(\epsilon)=(\epsilon I^{bb} - H^{bb}_0)^{-1}.
  \label{Rdef1}
  \end{equation}
The resolvent $R(\epsilon)$ is defined when
$\epsilon=\epsilon_{0k}$ for $k\in \Levels$, which is outside the range of
eigenvalues of $H^{bb}_0$.  Thus we can write
\begin{equation}
  P_1 = \left(\begin{array}{cc}
      0 & P^{ab}_1 \\
      P^{ba}_1 & 0
      \end{array}\right),
    \label{P1soln}
    \end{equation}
where $P^{ba}_1 = (P^{ab}_1)^\dagger$.

At second order we require $P_0P_2 + P_1^2 + P_2P_0=P_2$,
which gives $P^{aa}_2 = -P^{ab}_1\,P^{ba}_1$ and $P^{bb}_2 = +
P^{ba}_1 \, P^{ab}_1$.   We also require
$[P_0,H_2]+[P_1,H_1]+[P_2,H_0]=0$, which gives
\begin{equation}
  P^{ab}_{2,kl} = \sum_{m\notin \Levels}
  \big(H^{ab}_2 + P^{ab}_1\, H^{bb}_1
  -H^{aa}_1\,P^{ab}_1\big)_{km}\,R(\epsilon_{0k})_{ml}.
  \label{P2absoln}
  \end{equation}
Thus we can write
\begin{equation}
  P_2=\left(\begin{array}{cc}
      -P^{ab}_1 P^{ba}_1 & P^{ab}_2 \\
      P^{ba}_2 & P^{ba}_1 P^{ab}_1
    \end{array}\right).
  \label{P2soln}
  \end{equation}
Finally, at third order we find
\begin{equation}
  P_3=\left(\begin{array}{cc}
      -P^{ab}_1 P^{ba}_2-P^{ab}_2 P^{ba}_1 & P^{ab}_3 \\
      P^{ba}_3 & P^{ba}_1 P^{ab}_2 + P^{ba}_2 P^{ab}_1
      \end{array}\right)
    \label{P3soln}
    \end{equation}
where
\begin{eqnarray}
  P^{ab}_{3,kl}&=&\sum_{m\notin \Levels} (H^{ab}_3 + P^{ab}_1 H^{bb}_2
  -H^{aa}_2 P^{ab}_1 -P^{ab}_1 P^{ba}_1 H^{ab}_1\nonumber\\
  &+&P^{ab}_2 H^{bb}_1
  -H^{aa}_1 P^{ab}_2 -H^{ab}_1 P^{ba}_1 P^{ab}_1)_{km} R(\epsilon_{0k})_{ml},
  \end{eqnarray}
and $P^{ba}_3 = (P^{ab}_3)^\dagger$.

Now we can take the square roots indicated in (\ref{TintermsofP}),
which to third order are given by
\begin{subequations}
\begin{eqnarray}
  (P^{aa})^{1/2} &=& I^{aa} -\frac{1}{2} P^{ab}_1 P^{ba}_1
  -\frac{1}{2}(P^{ab}_1 P^{ba}_2 +P^{ab}_2P^{ba}_1), \\
  (Q^{bb})^{1/2} &=& I^{bb} -\frac{1}{2} P^{ba}_1 P^{ab}_1
  -\frac{1}{2}(P^{ba}_1 P^{ab}_2 +P^{ba}_2 P^{ab}_1).
  \end{eqnarray}
\end{subequations}
These allow us to write out the expansion of $T$,
\begin{widetext}
\begin{eqnarray}
  T&=&\left(\begin{array}{cc} I^{aa}& 0 \\ 0 & I^{bb}\end{array}\right)
    +\left(\begin{array}{cc} 0 & -P^{ab}_1 \\
        P^{ba}_1 & 0\end{array}\right)
    +\left(\begin{array}{cc} -(1/2)P^{ab}_1 P^{ba}_1 & -P^{ab}_2 \\
        P^{ba}_2 & -(1/2) P^{ba}_1 P^{ab}_1 \end{array}\right)
    \nonumber\\
    &+&\left(\begin{array}{cc} -(1/2)(P^{ab}_1 P^{ba}_2 
        +P^{ab}_2 P^{ba}_1) & -P^{ab}_3 -(1/2) P^{ab}_1 P^{ba}_1
        P^{ab}_1 \\
        P^{ba}_3 + (1/2) P^{ba}_1 P^{ab}_1 P^{ba}_1 &
        -(1/2)(P^{ba}_1 P^{ab}_2 + P^{ba}_2 P^{ab}_1)
        \end{array}\right),
      \label{Texpn}
    \end{eqnarray}
    \end{widetext}
which gives the expansion of the singular-value diabatic basis about
an arbitrary point $x_0$ to third order.

\section{A Diabatic Basis via Parallel Transport}
\label{diabaticparallel}

Parallel transport is a standard topic in gauge theories
(\cite{KobayashiNomizu63,Nakahara03}) and is an important part of the
theory of Berry's phases (\cite{Berry84,Simon83}).  Its role in
Born-Oppenheimer theory is also well known (\cite{Mead92,Bohmetal91}).
Parallel transport can be used to create a diabatic basis; one simply
takes a basis at a reference point $x_0$, possibly on a conical
intersection, and parallel transports it along radial lines emanating
from $x_0$ to fill in a basis defined over a neighborhood of $x_0$.
Such bases have a long history in the chemical literature, going back
at least to \textcite{Smith69}, who used parallel transport in
diatomics to create a strictly diabatic basis, possible in his case
because his parameter space was 1-dimensional.  Strictly diabatic
bases (by any means of construction) do not exist in the polyatomic
molecules considered in this article (\cite{MeadTruhlar82}), but for a contrary opinion see
\textcite{Baer75,Baer76,Baer80,BaerEnglman92,BaerAlijah00,Baer00,
Baer00a,Baer01,Baer02}.

\subsection{Parallel Transport}
\label{reviewpt}

We review some aspects of parallel transport that we will need; for a
slightly different perspective see \textcite{Mead92}.  Given
a vector $\psi(x) \in \Sspace(x)$, we wish to find the vector
$\psi(x+dx)=\psi(x)+d\psi$ at a nearby point such that $\psi(x+dx) \in
\Sspace(x+dx)$ and such the square distance in Hilbert space
\begin{equation}
  \big| \psi(x+dx) - \psi(x)\big|^2
  \label{psisqdist}
  \end{equation}
is minimum.  Writing simply $\psi$ for $\psi(x)$ and likewise $P$ for
$P(x)$, we have $P\psi=\psi$ and $(P+dP)(\psi+d\psi)=\psi+d\psi$, or,
\begin{equation}
  d\psi=Pd\psi + (dP)\psi.
  \label{dpsieqn}
  \end{equation}
We regard $\psi$ as given and $d\psi$ as unknown.  Since
$d\psi=Pd\psi+Qd\psi$ we see that $(dP)\psi=Qd\psi$ and that the two
terms in (\ref{dpsieqn}) are orthogonal.  Also, the term $(dP)\psi$,
the component of $d\psi$ orthogonal to $\Sspace(x)$, is fixed;
therefore to minimize the square distance (\ref{psisqdist}) we choose
$d\psi$ so that it is orthogonal to $\Sspace(x)$, that is, so that
$Pd\psi=0$.

This gives $d\psi=(dP)\psi$.  If now the step from $x$ to $x+dx$ is
taken along a curve $x(\lambda)$, then by dividing by $d\lambda$ we
obtain an equation of parallel transport along the curve,
\begin{equation}
  \psi' = P'\psi,
  \label{psitransport}
  \end{equation}
where the prime means $d/d\lambda$.  This equation is to be used only
for vectors $\psi(x)$ that lie in $\Sspace(x)$, that is, that satisfy
$Q\psi=0$.   It can be shown (see Appendix~\ref{proofs}) that 
(\ref{psitransport}) implies
\begin{equation}
  \left(\frac{d}{d\lambda}+P'\right)(Q\psi)=0,
  \label{Qpsieqn}
  \end{equation}
so that if $Q\psi=0$ at $\lambda=0$ then $Q\psi=0$ for all $\lambda$
(this follows from the uniqueness of the solution of the differential
equation). Similarly, if we wish to transport a
vector that remains in $\Sspace^\perp(x)$ if it starts out in  that
space, then we use the transport equation
\begin{equation} 
  \psi'=Q'\psi=-P'\psi.
  \label{psitransport2}
  \end{equation}

We can combine both types of parallel transport into a single equation
by writing
\begin{equation}
  \psi'=(P'P-PP')\psi = [P',P]\psi.
  \label{psitransport1}
\end{equation}
It can be shown that if $\psi$ satisfies this equation then $Q\psi$
satisfies (\ref{Qpsieqn}), so that if $\psi\in\Sspace(x)$ initially
then it remains in $\Sspace(x)$; and it can be shown that if $\psi$
satisfies this equation and $Q\psi=0$ then $\psi$ satisfies the
simpler equation (\ref{psitransport}). (See Appendix~\ref{proofs}.)
Thus, (\ref{psitransport}) and (\ref{psitransport1}) are equivalent if
$\psi\in\Sspace(x)$.  Similarly it can be shown that
(\ref{psitransport2}) and (\ref{psitransport1}) are equivalent if
$\psi\in\Sspace^\perp(x)$.

The form (\ref{psitransport1}) is convenient because it applies to the
transport of any vector, one in $\Sspace(x)$, one in
$\Sspace^\perp(x)$, or any linear combination thereof.  Also, since
the commutator $[P',P]$ is anti-Hermitian, it follows that parallel
transport via (\ref{psitransport1}) is unitary, and preserves the
scalar product of vectors.   In particular, the parallel transport of
an orthonormal frame is orthonormal.   Furthermore, if we parallel
transport a frame that block diagonalizes the Hamiltonian, it
continues to do so as we evolve along the curve.  Finally, we note
that another useful form for (\ref{psitransport1}) is
\begin{equation}
  \psi' = (P'P + Q'Q)\psi.
  \label{psitransport3}
  \end{equation}
We mention this because it makes it evident how parallel transport is generalized to the case in which there are multiple subspaces of the electronic Hilbert spaee under consideration (not just two).

\subsection{The Parallel Transported Basis}
\label{paralleltransportbasis}

Now let us replace the displacement $\xi$ by $\lambda\xi$, so that
\begin{equation}
	x^\mu(\lambda) = x_0^\mu + \lambda \xi^\mu,
        \label{radialline}
	\end{equation}
which represents a radial line parameterized by $\lambda$ starting at
$x_0$ when $\lambda=0$.  We expand the Hamiltonian $H(x)$ and
projector $P(x)$ as before, except now with a $\lambda$ dependence, so
that for example
\begin{equation}
	H(x(\lambda)) = H_0 + \lambda H_1 + \lambda^2 H_2
	+\lambda^3 H_3+\ldots,
	\end{equation}
instead of (\ref{Hexpand}).  We expand $P(x(\lambda))$ and a
parallel-transported wave function $\psi(\lambda)$ similarly.  The
wave function $\psi$ is initially defined only along a single curve,
so it is a function of $\lambda$ but not of $x$.

Then we have
\begin{subequations}
\label{Phierarchy}
\begin{eqnarray}
  P(x) &=& P_0 + \lambda P_1 + \lambda^2 P_2 + 
	\lambda^3 P_3 +\ldots,\\
  P'(x) &=& P_1 + 2\lambda P_2 + 3\lambda^2 P_3 +\ldots,\\
  \relax[P',P] &=& [P_1,P_0] + 2\lambda[P_2,P_0]
  \nonumber\\
	&+&\lambda^2 \big( 3[P_3,P_0]+[P_2,P_1]\big)+\ldots
  \end{eqnarray}
\end{subequations}
and with $\psi=\psi_0+\lambda\psi_1 + \lambda^2\psi_2 + \lambda^3
\psi_3 +\ldots$, (\ref{psitransport1}) becomes
\begin{subequations}
\label{psihierarchy}
\begin{eqnarray}
  \psi_1 &=& [P_1,P_0]\psi_0,\\
  2\psi_2 &=& [P_1,P_0]\psi_1 + 2[P_2,P_0]\psi_0,\\
  3\psi_3 &=& [P_1,P_0]\psi_2 + 2[P_2,P_0]\psi_1
  \nonumber\\
    &+&\big(3[P_3,P_0]+[P_2,P_1]\big)\psi_0,
  \end{eqnarray}
\end{subequations}
etc.  These allow us to express $\psi_1$, $\psi_2$, etc., in terms of
$\psi_0$.  

We write the result in the following way.  Define $\psi_k(\lambda)$ as
the components of $\psi(\lambda)$ with respect to the reference basis
$\ket{ax_0;k}$, that is, $\psi_k(\lambda) =\braket{ax_0;k}
{\psi(\lambda)}$.  Then $\psi_k(\lambda)$ is a linear function of
$\psi_k(0)$, which is expressed by
\begin{equation}
  \psi_k(\lambda)=\sum_{{\rm all}\;l} T_{p;kl}\, \psi_l(0),
  \label{Tpdef}
  \end{equation}
where $T_p$ is an infinite-dimensional matrix that depends on
$\lambda$ along a single curve, or, with $\lambda=1$ and $\xi$
regarded as a variable, on $x=x_0+\xi$.  This matrix plays a
similar role to $T$ for the singular-value diabatic basis and, in
particular, if we define $\ket{px;k}$ as the basis that is parallel
transported along radial lines from $x_0$, with the initial values
being the reference basis $\ket{ax_0;k}$ at $x_0$, then
\begin{equation}
  \ket{px;k} = \sum_{{\rm all}\;l} \ket{ax_0;l} [T_p(x)]_{lk},
  \label{ptbasisdef}
  \end{equation}
which may be compared to (\ref{Tdef}).  The subscript on $T_p$ refers
to the parallel-transported basis.

Now we expand $T_p = T_{p0}+\lambda T_{p1}+\lambda^2 T_{p2}+\ldots$
and use (\ref{psihierarchy}) to obtain
\begin{subequations}
\label{Tphierarchy}
\begin{eqnarray}
  T_{p1} &=& [P_1,P_0],\\
  T_{p2} &=& \frac{1}{2}[P_1,P_0] T_{p1} + [P_2,P_0],\\
  T_{p3} &=& \frac{1}{3}[P_1,P_0] T_{p2} + \frac{2}{3}
             [P_2,P_0]T_{p1}
    \nonumber\\           
             &+&[P_3,P_0] + \frac{1}{3}
             [P_2,P_1].
  \end{eqnarray}
\end{subequations}
Then we set $\lambda=1$ and work out the commutators to obtain $T_p$
through third order,
\begin{widetext}
\begin{eqnarray}
  T_p&=&\left(\begin{array}{cc} I^{aa}& 0 \\ 0 & I^{bb}\end{array}\right)
    +\left(\begin{array}{cc} 0 & -P^{ab}_1 \\
        P^{ba}_1 & 0\end{array}\right)
    +\left(\begin{array}{cc} -(1/2)P^{ab}_1 P^{ba}_1 & -P^{ab}_2 \\
        P^{ba}_2 & -(1/2) P^{ba}_1 P^{ab}_1 \end{array}\right)
    \nonumber\\
    &+&\left(\begin{array}{cc}
        -(2/3)P^{ab}_1 P^{ba}_2 -(1/3)P^{ab}_2 P^{ba}_1 &
        -P^{ab}_3 -(1/2)P^{ab}_1 P^{ba}_1 P^{ab}_1 \\
        P^{ba}_3 + (1/2)P^{ba}_1 P^{ab}_1 P^{ba}_1&
        -(2/3)P^{ba}_1 P^{ab}_2 -(1/3) P^{ba}_2 P^{ab}_1
        \end{array}\right).
      \label{Tpexpn}
    \end{eqnarray}
\end{widetext}
Comparing this with (\ref{Texpn}) we see that the singular-value
diabatic basis and the parallel-transported diabatic basis are
identical through second order in an expansion about $x_0$ but differ
at third order.  The two expansions cannot be identical to all orders,
since in effect each small step of the parallel transport involves an
infinitesimal version of the singular-value basis, but one that
compares frames between $x$ and $x+dx$ rather than between $x_0$ and
$x$.

\subsection{Hamiltonian in Parallel-Transported Basis}

Let us write
\begin{equation}
  H_{p,kl}(x)=\matrixelement{px;k}{H(x)}{px;l}
  \label{Hpkldef}
  \end{equation}
for the Hamiltonian matrix in the parallel-transported basis, while
other matrices ($H_1$, $P_2$, etc.) are understood to be in the
reference basis $\ket{ax_0;k}$, as previously.   Then
\begin{equation}
  H_p(x) = T_p(x)^\dagger H(x) T_p(x).
  \end{equation}
Now we expand $H(x)$ as in (\ref{Hexpand}) and $T_p(x)$ as in
(\ref{Tphierarchy}) and multiply series to compute $H_p(x)$ to second
order.  We find that the off-diagonal elements vanish,
$H^{ab}_p(x)=H^{ba}_p(x)=0$, as they should, while through first order
the diagonal elements of $H_p(x)$ are the same as those of $H(x)$,
that is, $H^{aa}_{p0}=H^{aa}_0$, $H^{aa}_{p1}=H^{aa}_1$,
$H^{bb}_{p0}=H^{bb}_0$ and $H^{bb}_{p1}=H^{bb}_1$.  As for the
diagonal elements at second order, we find
\begin{eqnarray}
  H^{aa}_{p2} &=& H^{aa}_2 
  +\frac{1}{2}(H^{ab}_1 P^{ba}_1 + P^{ab}_1 H^{ba}_1), \\
  H^{bb}_{p2} &=& H^{bb}_2 
  -\frac{1}{2}(H^{ba}_1 P^{ab}_1 + P^{ba}_1 H^{ab}_1).
  \end{eqnarray}
We note that if the electronic Hamiltonian at $x_0$ is completely
degenerate in the coupled subspace, then we can use (\ref{completelydegen})
to obtain
\begin{equation}
  H^{aa}_{p2} = H^{aa}_2 + H^{ab}_1 R(\epsilon_0) H^{ba}_1,
  \end{equation}
the usual result from degenerate perturbation theory.

Now by diagonalizing $H^{aa}_p$ we obtain the adiabatic basis
$\ket{ax;k}$, if we want it.  If $x_0$ is on an internal degeneracy (a
seam or conical intersection), then this will reveal the usual
singularities of the adiabatic basis, for example, the dependence of
$\ket{ax;k}$ on the direction of approach as $x\to x_0$ and the $\pi$
phase discontinuity as we encircle the degeneracy manifold in the case
of a 2-fold degeneracy in the electrostatic model
(\cite{HerzbergLonguetHiggins63, LonguetHiggins75, MeadTruhlar79}).  In
the case of a triatomic with an odd number of electrons and spin-orbit
effects included it will reveal the string of a Dirac monopole
(\cite{Mead80a, Mead87}).

\subsection{The Derivative Couplings and Connection}
\label{DerivativeCandConnection}

If $\ket{x;k}$ is any smooth basis that block diagonalizes the
Hamiltonian then we define the derivative couplings by
\begin{equation}
  F_{\mu,kl} = \matrixelement{x;k}{\partial_\mu}{x;l}
  =-(\partial_\mu\bra{x;k})\ket{x;l}=-F_{\mu,lk}^*,
  \label{Fmukldef}
  \end{equation}
where $\partial_\mu=\partial/\partial x^\mu$ and where the alternative
forms follow from the orthonormality of the basis vectors.  For
example, the basis $\ket{x;k}$ could be $\ket{dx;k}$ or $\ket{px;k}$
or the adiabatic basis $\ket{ax;k}$ in regions where the latter is
smooth.  If we just write $F_\mu$ we mean the infinite dimensional
matrix whose components are $F_{\mu,kl}$; this can be decomposed into
$aa$-, $ab$-, etc., blocks.  As is well known, the derivative
couplings 
appear when the molecular Schr\"odinger equation is transformed to the
Born-Oppenheimer representation.  The matrix of derivative couplings
$F_\mu$, as well as its diagonal blocks, $F^{aa}_\mu$ and
$F^{bb}_\mu$, are anti-Hermitian and thus belong to the Lie algebra of
the unitary group.

Any process of parallel transport gives rise to a {\it connection},
which in the general case is a rule for associating a unique object at
a point $x+dx$, given that object at point $x$.  The object in
question could be a vector such as $\psi$, a frame, or other things.
In the case of transporting the vector $\psi$ as in
Sec.~\ref{reviewpt}, the components of the connection, which we denote
by $\Gamma_{\mu,kl}$, are defined by
\begin{equation}
  \frac{d\psi_k}{d\lambda} = -\sum_{{\rm all}\,l} \Gamma_{\mu,kl} 
  \frac{dx^\mu}{d\lambda}\, \psi_l,
  \label{Gammadef}
  \end{equation}
where the minus sign is conventional.  If we now substitute 
\begin{equation}
  \ket{\psi(\lambda)} = \sum_{{\rm all}\,k} \ket{x(\lambda);k} 
  \,\psi_k(\lambda)
  \label{psikexpn}
  \end{equation}
into the equation of parallel transport (\ref{psitransport1}) or
(\ref{psitransport3}) we find (see Appendix~\ref{proofs}),
\begin{equation}
  \Gamma_\mu = \left(\begin{array}{cc}
      F^{aa}_\mu & 0 \\
      0 & F^{bb}_\mu
      \end{array}\right),
    \label{GammaF}
    \end{equation}
that is, the diagonal blocks of the connection are the same as the
derivative couplings.   The off-diagonal components of  the connection
vanish because of the way we designed the process of parallel
transport in Sec.~\ref{reviewpt}, that is, so that a vector that
begins in one of the subspaces $\Sspace(x)$ or $\Sspace^\perp(x)$
remains in that subspace.

If $\ket{bx;k}$ is any smooth basis that block-diagonalizes the
Hamiltonian, where $b$ just means ``basis,'' then another such basis
is given by
\begin{equation}
  \ket{b'x;k} = \sum_l \ket{bx;l}\,U_{lk}(x),
  \label{changeofbasis}
  \end{equation}
where $U(x)$ is unitary, a smooth function of $x$ and also 
block-diagonal.  Then the derivative couplings in the new basis are
given by
\begin{equation}
  F'_\mu = U^\dagger F_\mu U + U^\dagger \partial_\mu U,
  \label{Fxfm}
  \end{equation}
which, as is well known, is the transformation law for a (generally)
non-Abelian gauge potential.  Because of the inhomogeneous term the
derivative couplings can be made to take on any value at any given point
by means of such a gauge transformation.  Indeed, in both the
singular-value and parallel-transported basis the derivative couplings
vanish at the reference point $x_0$.  Since a common goal in practice
is to minimize the derivative couplings over some region, naturally
the question arises as to how fast these couplings grow (in some
basis) as we move away from a point where they are known to vanish.
This question leads us to the curvature, which we take up next.

\subsection{Curvature}

Let $\ket{bx;k}$ be an arbitrary, smooth basis that block-diagonalizes
the Hamiltonian, as above, and let $\xi^\mu$ and $\eta^\mu$ be two
infinitesimal displacements from a point $x$.  Then let us carry out a
parallel transport around the parallelogram, $x\to x+\xi \to
x+\xi+\eta \to x+\eta \to x$.  Initially we parallel transport the
$a$-part of the frame, that is, the vectors $\ket{bx;k}$ for $k\in \Levels$.
This is a standard calculation in gauge theories
(\cite{KobayashiNomizu63, Nakahara03, Mead92}) and we just quote the
results.  Since parallel transport preserves orthonormality, the
parallel-transported frame, once again back at the starting point $x$,
must be related to the original frame at that point by a unitary
matrix.  Moreover, the unitary matrix must be infinitesimal, that is,
the identity plus an infinitesimal, anti-Hermitian correction.  If we
denote the parallel-transported frame, back at $x$, by $\ket{{\tilde
b}x;k}$, then we have
\begin{equation}
  \ket{{\tilde b}x;k} = \sum_{l\in \Levels} \ket{bx;l}\, [\delta_{lk}-
  \xi^\mu \eta^\nu \, G^{aa}_{\mu\nu,lk}(x)],
  \label{Gtransport}
  \end{equation}
which defines the $aa$-block of the curvature tensor $G_{\mu\nu;kl}$
at $x$.  In this formula, $G$ is understood to be expressed in the
basis $\ket{bx;k}$; if we omit the $k,l$ indices, we refer to the
matrix $G^{aa}_{\mu\nu}$.   The calculation shows that
\begin{equation}
  G^{aa}_{\mu\nu} = \partial_\mu F^{aa}_\nu -\partial_\nu F^{aa}_\mu +
  [F^{aa}_\mu,F^{aa}_\nu].
  \label{Gdef}
\end{equation}
If we carry out this calculation for the $b$-block, we find the same
equation but with $aa$ replaced by $bb$; and the off-diagonal blocks
of $G$ vanish, $G^{ab}_{\mu\nu} = G^{ba}_{\mu\nu}=0$, for the same
reason that the off-diagonal blocks of the connection do (see
(\ref{GammaF})). 

Given the fact that $F^{aa}_\mu$ and $F^{bb}_\mu$ are anti-Hermitian, it
follows from (\ref{Gdef}) that $G^{aa}_{\mu\nu}$ and $G^{bb}_{\mu\nu}$
are, too, that is, $G^{aa}_{\mu\nu,kl}=-G^{aa*}_{\mu\nu,lk}$, etc.

The nuclear Hamiltonian involves the vector potential $F$, not the
curvature $G$, but the curvature appears in the commutators of the
velocity operators (\cite{LittlejohnReinsch97}) and in the classical
equations of motion, which are useful for wave packet evolution
(\cite{BerryRobbins93,WuMiaoSubotnik20,Bianetal21}), especially when
nuclear dynamics depend on spin.

We now present another formula for the curvature, which is a
simplification of Eq.~(3.42) of \textcite{Mead92} and which connects
the curvature with the projection operator.  It is
\begin{equation}
  G^{aa}_{\mu\nu,kl}(x) = \matrixelement{bx;k}{[P_\mu,P_\nu]}
  {bx;l},
  \label{GintermsofP}
  \end{equation}
where $P_\mu = \partial_\mu P$.  The proof of this is straightforward.
In our applications this turned out to be the most convenient means of
calculating the curvature (much more convenient than (\ref{Gdef})).

Under the change of basis (\ref{changeofbasis}) the curvature
transforms as a tensor, that is,
\begin{equation}
  G'_{\mu\nu} = U G_{\mu\nu} U^\dagger,
  \label{Gxfm}
  \end{equation}
(\cite{Mead92}), where $U$ is the same as in (\ref{Fxfm}).  This means
that the quantities
\begin{equation}
  \sum_{kl} |G_{\mu\nu,kl}|^2
  \label{Ginvariants}
  \end{equation}
are independent of basis.  These quantities, which constrain the
derivatives of the derivative couplings, cannot be transformed away,
not even at a single point.  For this reason, there is some interest
in finding a gauge that expresses $F_\mu$ in terms of $G_{\mu\nu}$. 

Finally, we quote the Bianchi identity, which is
\begin{eqnarray}
  &&\partial_\sigma G^{aa}_{\mu\nu} + \partial_\mu G^{aa}_{\nu\sigma}
  +\partial_\nu G^{aa}_{\sigma\mu}
    \nonumber\\ 
  &+&[F^{aa}_\mu,G^{aa}_{\nu\sigma}]
  +[F^{aa}_\nu,G^{aa}_{\sigma\mu}]+[F^{aa}_\sigma,G^{aa}_{\mu\nu}]
  =0,
  \label{bianchi}
\end{eqnarray}
with a similar formula for the $bb$-block.   This is the analog of
Maxwell's equation $\nabla\cdot\Bvec=0$ in electromagnetism.  

\subsection{Non-Abelian Poincar\'e Gauge} 

In electromagnetism, Poincar\'e gauge (\cite{CohenTannoudjietal89}) is
transverse in real space, that is, $\xvec\cdot\Avec(\xvec)=0$.  In a molecule, if
$x_0$ is a reference point in nuclear configuration space and
$x(\lambda)=x_0+\lambda\xi$ is a radial line, as above, then the
analog of Poincar\'e gauge for the Mead-Truhlar-Berry connection is a
gauge such that
\begin{equation}
  \xi^\mu F^{aa}_\mu\big(x(\lambda)\big)
  =\xi^\mu F^{bb}_\mu\big(x(\lambda)\big)=0,
  \label{poincare}
  \end{equation}
that is, the radial component of the connection vanishes.

It turns out that parallel-transported gauge is Poincar\'e gauge,
in this sense.  To prove this we start with $k,l\in \Levels$ and note that
in the parallel-transported frame we have
\begin{eqnarray}
  \xi^\mu F^{aa}_{\mu,kl}&=&\xi^\mu \matrixelement{px;k}{\partial_\mu}
  {px;l} = \matrixelement{px;k}{\frac{d}{d\lambda}}{px;l}
  \nonumber\\
  &=&\matrixelement{px;k}{[P',P]}{px;l}
    =\matrixelement{px;k}{P(P'P-PP')P}{px;l}
    \nonumber\\
    &=&0,
  \end{eqnarray}
where we use $d/d\lambda=\xi^\mu\partial_\mu$ and
(\ref{psitransport1}).  A similar proof holds for $F^{bb}_\mu$, but
radial components of the off-diagonal blocks of the derivative
couplings, $F^{ab}_\mu$ and $F^{ba}_\mu$, do not vanish.

Poincar\'e gauge has the interesting property that the vector
potential can be expressed in terms of the magnetic field.   Thus one
obtains Hamiltonians that are not expressed in terms of some arbitrary
vector potential, but rather directly in terms of the magnetic field,
which is fully physical.   This is mainly useful in some neighborhood
of a given point, by means of Taylor series expansions.   

This property carries over to the non-Abelian case.   Specifically, if
$F^{aa}_\mu$ is in parallel-transported gauge, then
\begin{equation}
  F^{aa}_{\mu}(x) = \int_0^1 d\lambda\, \lambda\xi^\nu G^{aa}_{\nu\mu}
  \big(x(\lambda)\big),
  \label{FGeqn}
  \end{equation}
where $x$ on the left-hand side means $x=x_0+\xi$ and $x(\lambda)$
under the integral means $x_0+\lambda\xi$, as above.  The integral is
carried along a radial line from $x_0$ to $x=x_0+\xi$.  A similar
equation holds for the $bb$-block of the connection and curvature.

This is a rather remarkable formula, because it appears to give a
linear relationship between $F$ and $G$, which in the non-Abelian case
are in fact connected by the nonlinear relationship (\ref{Gdef}).  For
us (\ref{FGeqn}) is merely an identity connecting the curvature and
connection in parallel-transported gauge, but if one were to interpret
it as a map from a given curvature to a corresponding vector
potential, then it must be understood that the formula only works if
$G$ satisfies the Bianchi identity (\ref{bianchi}).

To prove (\ref{FGeqn}) we assume that $F$ is the parallel-transported
connection and $G$ is the curvature and we examine  the integral on
the right-hand side, but without assuming that it equals the left-hand
side.  Writing $G$ in terms of $F$, the integral becomes
\begin{equation}
  \int_0^1 d\lambda \, \lambda\xi^\nu \big(\partial_\nu F^{aa}_\mu
  -\partial_\mu F^{aa}_\nu + [F^{aa}_\nu,F^{aa}_\mu]\big),
  \label{FGintegral}
  \end{equation}
where it is understood that everything under the integral is evaluated
at $x(\lambda)=x_0+\lambda\xi$.  Because $\xi^\nu
F^{aa}_\nu\big(x(\lambda)\big)=0$ the commutator term vanishes.  The
first term in (\ref{FGintegral}) can be written
\begin{equation}
  \int_0^1 d\lambda \, \lambda \frac{dF^{aa}_\mu}{d\lambda},
  \label{FGintegralT1}
  \end{equation}
where we use $d/d\lambda = \xi^\nu\partial_\nu$.  As for the second
term, it is
\begin{eqnarray}
  -\int_0^1 d\lambda \, \xi^\nu \frac{\partial F^{aa}_\nu}
  {\partial\xi^\mu}&=&
  -\int_0^1 d\lambda \left[\frac{\partial}{\partial\xi^\mu}
    (\xi^\nu F^{aa}_\nu) - F^{aa}_\mu\right]
    \nonumber\\
  &=&\int_0^1 d\lambda \, F^{aa}_\mu,
  \label{FGintegralT2}
  \end{eqnarray}
where we use $\partial/\partial\xi^\mu = \lambda\partial_\mu$.  Now
adding (\ref{FGintegralT1}) and (\ref{FGintegralT2}) we obtain
\begin{equation}
  \int_0^1 d\lambda \, \frac{d}{d\lambda}[\lambda 
  F^{aa}_\mu\big(x(\lambda)\big)] = F^{aa}_\mu\big(x(1)\big)
  =F^{aa}_\mu(x),
  \end{equation}
which is the given left-hand side of (\ref{FGeqn}). 

Now we may expand the integrand in (\ref{FGeqn}) in a power series in
$\lambda$ and do the integral, which through order $\lambda^2$ gives
us
\begin{equation}
  F^{aa}_\mu(x_0+\xi) = \frac{1}{2} \xi^\nu G^{aa}_{\nu\mu}(x_0)
  + \frac{1}{3}\xi^\nu \xi^\sigma (\partial_\sigma G^{aa}_{\nu\mu})
  (x_0) + \ldots
  \label{Fexpansion}
  \end{equation}
This can be used to express $F$ and its derivatives at $x_0$ in terms
of the curvature,
\begin{subequations}
\label{FGexpns}
\begin{eqnarray}
  F^{aa}_\mu &=& 0,\\
  \partial_\nu F^{aa}_\mu &=& \frac{1}{2} G^{aa}_{\nu\mu},\\
  \partial_\nu \partial_\sigma F^{aa}_\mu &=&
  \frac{1}{3}[\partial_\sigma G^{aa}_{\nu\mu} +
  \partial_\nu G^{aa}_{\sigma\mu}],
  \end{eqnarray}
\end{subequations}
where it is understood that everything is evaluated at $x_0$.  As we
see, the Mead-Truhlar-Berry connection in the parallel-transported
frame vanishes at $x_0$ and the curvature governs the rate at which it
grows as we move away from $x_0$.

The facility of expressing the connection in  terms of the curvature
is one of the advantages of the parallel-transported frame over the
singular-value frame. 

\subsection{Connection and Curvature in Parallel-Transported Basis}

We can now work out the components of the connection and curvature in
the parallel-transported basis.  It is  straightforward to do this 
with the expansion of the parallel-transported basis $\ket{px;k}$ in
terms of the reference basis $\ket{ax_0;k}$, given by
(\ref{ptbasisdef}), with coefficients given by (\ref{Tpexpn}).  First
we substitute this into the definition (\ref{Fmukldef}) of  the
derivative couplings to compute $F_\mu$, and then use (\ref{Gdef}) to
compute $G_{\mu\nu}$.   This will give expressions for $F_\mu$ and
$G_{\mu\nu}$ in  terms of the components of the projection operator,
and, ultimately, in terms of the Hamiltonian and its derivatives.

It turns out to be much easier, however, to use first Mead's formula
(\ref{GintermsofP}) in the parallel-transported basis to find $G$, and
then to use (\ref{FGeqn}) to find $F$. In this work we must be
careful with notation.  We will henceforth write $\Proj$ for the
projection operator (previously simply denoted $P$), $P$ for its
matrix with respect to the reference basis $\ket{ax_0;k}$ and $\Pbar$
for its matrix with respect to the parallel-transported basis.  Also,
we will append $(x)$ to a quantity if we mean it to be evaluated at
$x=x_0+\xi$, and we will omit the $(x)$ if we mean it to be evaluated
at $x_0$.  Also, we define $\Proj_\mu(x)=(\partial_\mu\Proj)(x)$, etc.

We are interested in the curvature tensor in the parallel-transported
basis, which according to Mead's formula (\ref{GintermsofP}) in the
case $k,l\in \Levels$ is
\begin{equation}
  G^{aa}_{\mu\nu;kl}(x) = \matrixelement{px;k}
  {[\Proj_\mu(x),\Proj_\nu(x)]}
  {px;l} = [\Pbar_\mu(x),\Pbar_\nu(x)]_{kl},
  \end{equation}
since the matrix of the commutator is the commutator of the matrices.
A similar formula applies to the $bb$-block. For this we need the
matrix,
\begin{equation}
  \Pbar_\mu(x) = T_p(x)^\dagger P_\mu(x) T_p(x),
  \label{Pbasischange}
  \end{equation}
which is a change of basis from $\ket{ax_0;k}$ to $\ket{px;k}$.  This
matrix vanishes on the diagonal blocks, for example, if $k,l \in \Levels$
then
\begin{eqnarray}
  [\Pbar_\mu(x)]_{kl} &=& \matrixelement{px;k}{\Proj_\mu(x)}
  {px;l}
  \nonumber\\
  &=& \matrixelement{px;k}{\Proj(x)\Proj_\mu(x)\Proj(x)}
  {px;l} =0,
  \end{eqnarray}
as follows since $\Proj(x)[\partial_\mu\Proj(x)]\Proj(x)=0$.

The expansion of $P(x)=P(x_0+\xi)$ is given by (\ref{PH0defs}),
(\ref{P1soln}), (\ref{P2soln}) and (\ref{P3soln}), which to second
order can be written
\begin{eqnarray}
  P(x)&=&\left(\begin{array}{cc} I^{aa} & 0 \\
      0 & 0\end{array}\right) +
  \left(\begin{array}{cc} 0 & P^{ab}_\mu \\
      P^{ba}_\mu & 0\end{array}\right)\xi^\mu
      \nonumber\\
  &+&\left(\begin{array}{cc} -P^{ab}_\mu P^{ba}_\nu &
      (1/2)P^{ab}_{\mu\nu} \\
      (1/2)P^{ba}_{\mu\nu} & P^{ba}_\mu P^{ab}_\nu
      \end{array}\right)\xi^\mu\xi^\nu + \ldots,
    \label{Pxiexpn}
    \end{eqnarray}
where $P^{ab}_\mu$ is the $ab$-block of $(\partial_\mu P)(x_0)$,
$P^{ab}_{\mu\nu}$ is the $ab$-block of $(\partial_\mu\partial_\nu P)
(x_0)$, etc.  Notice that in terms of our previous notation we have
\begin{equation}
  P^{ab}_1 = P^{ab}_\mu \,\xi^\mu, \qquad
  P^{ab}_2 = \frac{1}{2} P^{ab}_{\mu\nu}\,\xi^\mu\xi^\nu,
  \label{Pxiderivs}
  \end{equation}
etc.  Differentiating (\ref{Pxiexpn}) we obtain
\begin{eqnarray}
  &&P_\mu(x) = \left(\begin{array}{cc}
      0 & P^{ab}_\mu \\
      P^{ba}_\mu & 0\end{array}\right)
      \nonumber\\
  &+&\left(\begin{array}{cc}
      -P^{ab}_\mu P^{ba}_\nu - P^{ab}_\nu P^{ba}_\mu & 
      P^{ab}_{\mu\nu}\\
      P^{ba}_{\mu\nu} & P^{ba}_\mu P^{ab}_\nu + P^{ba}_\nu P^{ab}_\mu
      \end{array}\right)\xi^\nu + \ldots
    \end{eqnarray}
Now we use the expansion (\ref{Tpexpn}) of $T_p$, which we only need
to first order and which we write as
\begin{equation}
  T_p(x) = \left(\begin{array}{cc} I^{aa} & 0 \\ 0 & I^{bb}
      \end{array}\right) + \left(\begin{array}{cc}
        0 & -P^{ab}_\nu \\ P^{ba}_\nu & 0
        \end{array}\right)\xi^\nu + \ldots,
      \label{Tpexpn1}
      \end{equation}
to carry out  the change of basis (\ref{Pbasischange}).   This gives
\begin{equation}
  \Pbar_\mu(x) = \left(\begin{array}{cc}
      0 & P^{ab}_\mu \\ P^{ba}_\mu & 0
      \end{array}\right) +
    \left(\begin{array}{cc}
        0 & P^{ab}_{\mu\nu} \\
        P^{ba}_{\mu\nu} & 0
        \end{array}\right)\xi^\nu + \ldots
      \label{Pbarmuexpn}
      \end{equation}
which is off-diagonal, as predicted.  

Now we can compute the commutator and thence the curvature.  We find
\begin{equation}
  G^{aa}_{\mu\nu}(x) = (P^{ab}_\mu P^{ba}_\nu 
  -{\rm h.c.}) +(P^{ab}_\mu P^{ba}_{\nu\sigma} +
  P^{ab}_{\mu\sigma} P^{ba}_\nu - {\rm h.c.})\xi^\sigma + \ldots,
  \label{Gaaexpn}
  \end{equation}
where subtracting the Hermitian conjugate of the indicated matrices is
equivalent to antisymmetrizing in $\mu,\nu$.  The expression for
$G^{bb}(x)$ is obtained from this by swapping $a\leftrightarrow b$,
and the off-diagonal blocks vanish, $G^{ab}=G^{ba}=0$.  Finally,
replacing $\xi$ by $\lambda \xi$ and carrying out the integral
(\ref{FGeqn}), we obtain the diagonal blocks of the Mead-Truhlar-Berry connection or
derivative couplings in the parallel-transported basis.  We find
\begin{eqnarray}
  &&F^{aa}_{\mu}(x) = \frac{1}{2} (P^{ab}_\nu P^{ba}_\mu 
  -{\rm h.c.})\xi^\nu \nonumber\\
  &&+\frac{1}{6}
  (P^{ab}_\nu P^{ba}_{\mu\sigma}+P^{ab}_\sigma P^{ba}_{\mu\nu}
  +2P^{ab}_{\nu\sigma} P^{ba}_\mu
  -{\rm h.c.})
  \xi^\nu \xi^\sigma + \ldots,
  \label{FintermsofP}
  \end{eqnarray}
and similarly for the $bb$-block.

\subsection{Coordinates and Rotations}
So far we have been using the notation $x$ to stand for a point of the
nuclear configuration space in the center of mass frame and $x^\mu$
for the collection of the $3N-3$ components of the $N-1$ Jacobi
vectors.  We will call these ``Jacobi coordinates.''  Thus derivatives
$\partial_\mu$ are understood to be taken with respect to Jacobi
coordinates.  However, there is nothing in our analysis so far that
has required that the coordinates be these; everything goes through
with an arbitrary coordinate system on nuclear configuration space,
that is, a set of $3N-3$ arbitrary, possibly nonlinear functions
$x^{\prime\,\mu}$ of $x^\mu$.  We only require that the Jacobian
matrix $\partial x^{\prime\,\mu}/\partial x^\nu$ be nonsingular in the
neighborhood in which we are working.  Therefore all our formulas so
far are valid with $x^\mu$ reinterpreted as an arbitrary coordinate
system in this sense.

Our parallel-transported basis has been defined by integrating the
(\ref{psitransport1}) along radial lines emanating from a reference
point $x_0$, which have the coordinate representation
(\ref{radialline}).  This means that the parallel-transported basis at
a point $x$ depends on the coordinates, since a line that is straight
in one coordinate system is not straight in another.

\begin{figure}
\includegraphics[scale=0.5]{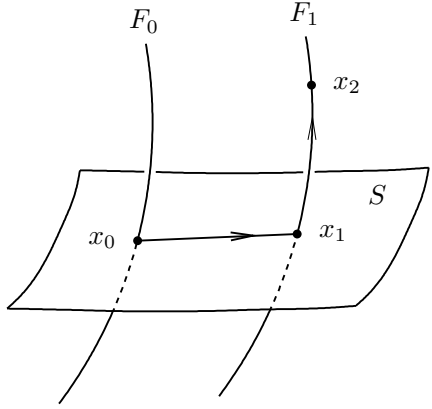}%
\caption{\label{section}Fibers $F_0$, $F_1$ are 3-dimensional surfaces
  swept out by applying rotations to nuclear configurations $x_0$,
  $x_1$.  Section $S$ is a $(3N-6)$-dimensional surface transverse to
  the fibers.  To establish a frame of electronic wave functions over
  a region of nuclear configuration space, we may use parallel
  transport along a section $S$, such as along the path $x_0\to x_1$,
  and then rotation operators to transport the frame along the fibers,
  such as along the path $x_1\to x_2$.}%
\end{figure}

Another issue concerns the direction of integration.   Let us assume
that the configuration $x_0$ is noncollinear, the typical situation in
polyatomic systems.   Then there are  three directions in which one
can move away from $x_0$ that are purely rotational, and we do not
want to use parallel transport to create a diabatic basis in  those
directions.  In those directions the nuclear configuration transforms
by a rigid rotation, that is, the shape of the molecule does not
change, only its orientation.   Those directions are tangent to the
the 3-dimensional surfaces which are generated by applying rotations
to a given configuration, which are the orbits or fibers of the action
of rotations on the nuclear configuration space, as explained by
\textcite{LittlejohnReinsch97}.   When the nuclear configuration
changes by a rigid rotation, the electronic basis functions should
change by a rotation operator, not by parallel transport (or any other
method for constructing a diabatic basis, such as the singular value
method).   Attention to those directions that are purely rotational versus those in which the shape changes is important in the construction of kinetic energy operators on the internal space (\cite{WangCarrington00}).

The situation is illustrated in Fig.~\ref{section}, which is a
schematic illustration of a region of the nuclear configuration
space.  Configuration $x_0$ is some configuration, and $F_0$ is the
set of other configurations related to $x_0$ by rigid rotations, that
is, configurations of the same shape as $x_0$ but different
orientations.  Surface $F_0$ is called the ``fiber through $x_0$'', as
explained by \textcite{LittlejohnReinsch97}.   Similarly $F_1$ is the
fiber through $x_1$.   Configurations $x_1$ and $x_2$ have the same
shape but different orientations.

Surface $S$ is a $(3N-6)$-dimensional surface that cuts transversally
through the fibers in the given region, called a ``section'' of the
fiber bundle (\cite{LittlejohnReinsch97}).  Sections are needed to
define orientational coordinates, that is, the section is the surface
upon which the Euler angles are those of the identity rotation.   The
section also defines a body frame; it is the surface upon which the
body frame coincides with the space frame.   Sections are also
implicitly used in electronic structure calculations, when it is
attempted to fill in some region of nuclear configuration space.
This is because one need only calculate the energy eigenvalues and
eigenfunctions for a single orientation of a given shape; for other
orientations the energy eigenvalues do not change and the energy
eigenfunctions change simply by a rotation operator.  Thus, electronic
structure calculations take place on a section.  

Now assume that $x_0$ is a reference configuration and we have an
electronic basis at $x_0$ that block-diagonalizes the Hamiltonian, and
suppose we wish to find a diabatic basis at neighboring points such as
$x_2$ in the figure.   Then the procedure is to use some rule
(singular value, parallel transport, etc.) to create a diabatic basis
at $x_1$ on the section, and then to use rotation operators to create
a diabatic basis at $x_2$.

Thus we see that when we parallel transport a basis from a
configuration $x_0$, we do not need to go out in all $3N-3$ directions
in nuclear configuration space, but rather only along the $3N-6$
directions of a section passing through $x_0$.  So far we
have been interpreting the coordinates $x^\mu$ as $3N-3$ coordinates
on nuclear configuration space, but all our results hold if we
reinterpret them as $3N-6$ coordinates on a section, and we will
henceforth do this.  These coordinates can be taken to be shape
coordinates,  that is, quantities such as bond lengths and angles that
are invariant under overall rotations of the molecule.  Similar
considerations apply to all other methods (singular value, etc.) of
constructing a diabatic basis in the neighborhood of $x_0$; the
purely rotational directions are handled by rotation operators.

\section{Degeneracy Manifolds}
\label{degeneracymanifolds}

In the following it is assumed that we are working on a section 
with coordinates $x^\mu$, $\mu=1,\ldots,3N-6$, or some subset of it
which for simplicity we will call $S$.   We assume the set $\Levels$
(see (\ref{Idef})) contains at least two levels, so that internal
degeneracies are possible.  We define the {\it degeneracy manifold} as
the subset of $S$ upon which there is an internal degeneracy.
Degeneracy manifolds are usually called ``seams'' but we prefer the
alternative terminology as it is more descriptive.  Degeneracy
manifolds are usually surfaces of conical intersections, but it should
be noted that while all intersections of potential energy surfaces are
degeneracies they are not always conical.  We now address the
construction of a parallel-transported basis in a neighborhood of
the degeneracy manifold. 

\subsection{Models of the Electronic Hamiltonian}

Up to this point we have been assuming that the electronic Hamiltonian
was taken in the electrostatic model, but now we will address the more
general case in which fine structure effects are included.  We shall
assume that no external fields are acting so that the molecular
Hamiltonian has time-reversal symmetry.  This has an important effect
in the case of an odd number of electrons, in which the electronic
eigenstates become Kramers doublets.  Let us agree to call a ``level''
a single Kramers doublet in the case of an odd number of electrons,
which means the degeneracy is twice the number of levels.  In other
cases (electrostatic model or fine structure with an even number of
electrons) the degeneracy is equal to the number of levels.

With this understanding, we now review the simplest case of two
levels. If there is no symmetry in addition to time-reversal, the
codimension of the degeneracy manifold is 2 in the electrostatic model
or when fine structure is included and the number of electrons is even
(\cite{VonNeumannWigner29}), and it is 5 if fine structure is included
and the number of electrons is odd (\cite{Mead80a}).  If there are
symmetries then the picture is more complicated since in general the
symmetry only holds on submanifolds of configuration space, but an
important exception is the $C_s$ symmetry of triatomic molecules
(reflection in the plane of the molecule), which is global.  In the
case of triatomics with fine structure and an odd number of electrons,
the codimension of the degeneracy manifold is 3 (\cite{Mead80a}).  We
should also add that in the case of the electrostatic model we are
assuming that one is working within a given subspace of spin states,
that is, energy eigenstates of fixed $S^2$ and $S_z$, since
degeneracies between states of different spin have codimension 1,
whereas when fine structure is included we must enlarge the electronic
Hilbert space to include all spin states.

Let us denote the degeneracy manifold by $D$; its codimension is
counted inside the section $S$.  For example, in triatomics $S$ is
3-dimensional.  Then in the electrostatic model or with fine structure
and an even number of electrons, $D$ has codimension 2 or dimension 1,
that is, it is a curve in $S$; while if fine structure is included and
number of electrons is odd, then the codimension of $D$ is 3 and its
dimension is 0, that is, $D$ consists of isolated points inside $S$.
These codimension counts apply at generic points of configuration
space, where the Jacobian of the map from configuration space to
Hamiltonian matrix space is of maximal rank; where this is not the
case other phenomena arise, such as the bifurcation of degeneracy
manifolds (\cite{Yarkony00}).

\subsection{Reference point $x_0$ on Degeneracy Manifold}

Henceforth we will assume that the reference point $x_0$ for the
construction of a parallel-transported basis lies on $D$.  We will
also assume that at $x_0$ the electronic Hamiltonian is completely
degenerate in the coupled subspace, so that $H^{aa}_{0,kl} =
\epsilon_0 \,\delta_{kl}$, where $\epsilon_0$ is the degenerate
eigenvalue.  In that case we note that $P^{ab}_n$ can be expressed
purely in terms of matrix products,
\begin{subequations}
\label{completelydegen}
\begin{eqnarray}
  P^{ab}_1 &=& H^{ab}_1 R(\epsilon_0),\\
  P^{ab}_2 &=& [H^{ab}_2 + H^{ab}_1 R(\epsilon_0) H^{bb}_1
  -H^{aa}_1 H^{ab}_1 R(\epsilon_0)]R(\epsilon_0),
  \end{eqnarray}
\end{subequations}
etc., with $P^{ba}_n = (P^{ab}_n)^\dagger$.  These are simplified
versions of (\ref{P1absoln}) and (\ref{P2absoln}).

Let us also write the expansion of the Hamiltonian at $x=x_0+
\xi$ as a power series in $\xi$, just as we did for the projection
operator in (\ref{Pxiderivs}).  We define, for example, $H^{ab}_\mu =
(\partial_\mu H^{ab})(x_0)$, $H^{ab}_{\mu\nu} =
(\partial_\mu\partial_\nu H^{ab})(x_0)$, etc., so that
\begin{equation}
  H^{ab}_1 = H^{ab}_\mu \, \xi^\mu, \qquad
  H^{ab}_2 = \frac{1}{2}H^{ab}_{\mu\nu}\,\xi^\mu\xi^\nu,
  \end{equation}
etc., and similarly for the other blocks.  Then with the aid of
(\ref{completelydegen}) the derivatives of $P$ can be expressed in terms
of the derivatives of $H$, 
\begin{subequations}
  \label{PintermsofH}
  \begin{eqnarray}
  P^{ab}_\mu &=& H^{ab}_\mu R(\epsilon_0),\\
  P^{ab}_{\mu\nu} &=& [H^{ab}_{\mu\nu} +H^{ab}_\mu R(\epsilon_0)
      H^{bb}_\nu + H^{ab}_\nu R(\epsilon_0) H^{bb}_\mu
      \nonumber\\
     &&\quad -(H^{aa}_\mu H^{ab}_\nu 
     +H^{aa}_\nu H^{ab}_\mu) R(\epsilon_0)] R(\epsilon_0),
   \end{eqnarray}
\end{subequations}
etc.  

These can then be used to express the curvature and connection
in terms of the derivatives of the Hamiltonian.   For example,
(\ref{GintermsofP}) gives us the curvature, which becomes
\begin{equation}
  G^{aa}_{\mu\nu} = [H^{ab}_\mu R(\epsilon_0)^2 H^{ba}_\nu -
  {\rm h.c.}]+\ldots
  \label{GintermsofH}
  \end{equation}
Here we omit the first order correction as
it is rather lengthy.  As for  the connection, in the
parallel-transported basis $\ket{px;k}$ it is given in terms of the
curvature by (\ref{Fexpansion}) or (\ref{FGexpns}); thus, to first
order we have
\begin{equation}
  F^{aa}_\mu(x)=\frac{1}{2}[H^{ab}_\nu R(\epsilon_0)^2 H^{ba}_\mu
  -{\rm h.c.}]\xi^\nu + \ldots.
  \label{FintermsofH}
  \end{equation}

The result (\ref{GintermsofH}) agrees with Eq.~(56) of
\textcite{MeadTruhlar82}, in spite of the fact that those authors were
not using the parallel-transported basis.  The reason is that the
curvature is a tensor, and has the same form in all frames.   The same
cannot be said for the connection.  We believe the first order
correction to the curvature seen in (\ref{GintermsofP}) is new. 

\subsection{Diabatic Basis in a Neighborhood of the Degeneracy Manifold}

We now construct a version of the parallel-transported basis in a
neighborhood of the degeneracy manifold.   So far we have constructed
such a basis in the neighborhood of an arbitrary point $x_0$; we are
free to choose $x_0$ to lie on $D$, but we still have just a small
neighborhood of a single point, where Taylor series expansions are
valid.  We desire a construction that is valid over all of $D$, or at
least the parts of $D$ that form a smooth manifold.   (Places where $D$
bifurcates require a separate analysis, for which see
\textcite{Yarkony00}).  For example, in triatomic molecules where $D$ 
is a line inside the 3-dimensional section $S$, we desire a
parallel-transported basis that is valid in a tubular neighborhood of
$D$, where the two potential energy surfaces are strongly coupled.

\begin{figure}
\includegraphics[scale=0.5]{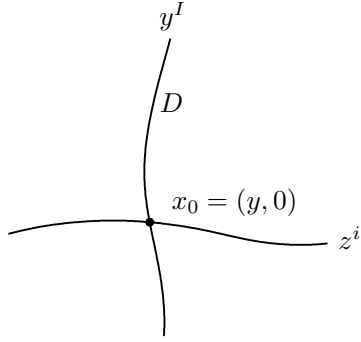}%
\caption{\label{yzcoords}The degeneracy manifold $D$ as a subset of
  the section $S$ and a neighborhood thereof are described by
  coordinates $x^\mu=(y^I,z^i)$, where $y^I$ are coordinates along $D$
  and $z^i$ are coordinates transverse to $D$.  Coordinates $z^i=0$ on
  $D$ itself.}%
\end{figure}

We let $x^\mu$, $\mu=1,\ldots,3N-6$ be coordinates on a section $S$,
which we break up as $x^\mu=(y^I,z^i)$, where $I=1,\ldots,\dim D$ and
$i=1,\ldots,\codim D$, making $y^I$ coordinates along the degeneracy
manifold, and $z^i$ coordinates transverse to it.  We let $z^i=0$ on
$D$, and we use the summation convention on coordinate indices $I$ and
$i$.  See Fig.~\ref{yzcoords}.  This is merely a coordinate
description of a neighborhood of $D$.  The whole construction takes
place in $S$, a subset of the nuclear configuration space which in
general is curved; and $D$ as a subset of $S$ is generally a curved
manifold.

These coordinates draw attention to the manifolds $y=y_0={\rm const.}$
illustrated in the figure which are transverse to $D$ and upon which
$z^i$ are coordinates.  These manifolds correspond roughly to what has
been called the ``branching space'' by \textcite{Yarkony04b}; there is
one such manifold for each point on $D$.  We propose to establish a
diabatic basis in a neighborhood of $D$ by parallel transporting along
radial lines in the transverse spaces away from $D$.  We will denote
this basis by $\ket{Dx;k}$; it is like the basis $\ket{px;k}$ except
that the reference point for the parallel transport, instead of being
fixed, moves along $D$ as $x$ moves.  That is, if $x=(y,z)$, then
for the basis $\ket{Dx;k}$ the reference point for the parallel
transport is $x_0=(y,0)$.

This construction requires that a diabatic basis be established on
$D$, before being propagated along the transverse spaces.  This can be
done by any convenient means, for example, the singular-value method
or by parallel transport.  Once this is done, not only do we have
frames for points $x_0=(y,0)\in D$, but the $y^I$-components of
the connection, that is,
\begin{equation}
	F^{aa}_{I;kl}(x_0)=\matrixelement{Dx_0;k}{\partial_I}{Dx_0;l},
	\label{FsubAonD}
	\end{equation}
where $\partial_I=\partial/\partial y^I$, become known at points $x_0\in
D$.

Once the basis $\ket{Dx_0;k}$ is established for $x_0=(y,0)\in D$,
we use parallel transport along radial lines in the transverse spaces
to create $\ket{Dx;k}$ in a neighborhood of $D$, that is, for
$x=(y,z)$.  The radial line is $x(\lambda)=(y,\lambda z)$, which
interpolates between $(y,0)$ and $(y,z)$ as $\lambda$ goes from 0
to 1.  The radial component of the connection in this basis
vanishes, for example, for $k,l\in \Levels$ we have
\begin{equation}
    z^i \matrixelement{Dx;k}{\partial_i}{Dx;l} =
    z^i F^{aa}_{i;kl}(y,z)=0,
    \label{Fitransverse}
    \end{equation}
which is like (\ref{poincare}) and proved in the same way.

\subsection{Connection and Curvature in a Neighborhood of $D$}
 
To find the connection and curvature in the basis $\ket{Dx;k}$ we
begin with the curvature.  Since the curvature is a tensor, the
numerical values of its components at a point $x$, say,
$G^{aa}_{\mu\nu;kl}(x)$, depend on the the coordinates and the
basis at $x$.  We wish to find these components in the basis
$\ket{Dx;k}$ where $x=(y,z)$.  But the basis $\ket{Dx;k}$ is the same
as the parallel-transported basis $\ket{px;k}$ if the reference point
for the latter is taken to be $x_0=(y,0)$.  So with that understanding
about the reference point, our expressions for the curvature in the basis
$\ket{px;k}$ are valid also in the basis $\ket{Dx;k}$.  We will,
however, want to break the coordinate indices up into their $y$- and
$z$-parts, by replacing indices $\mu$, etc., by $I$ or $i$, etc.   For
example, we can rewrite (\ref{Gaaexpn}) as
\begin{equation}
  G^{aa}_{ij}(y,z) = (P^{ab}_i P^{ba}_j -{\rm h.c.})
  +(P^{ab}_i P^{ba}_{jk} + P^{ab}_{ik} P^{ba}_j -{\rm h.c.})z^k 
  +\ldots,
  \label{Gaaijexpn}
  \end{equation}
and similarly for the $(iI)$, $(Ii)$ and $(IJ)$ components of $G$.
Here we have replaced $\xi^\mu$ by $(0,z^i)$, since the parallel
transport is taking place only in the transverse manifold (at constant
$y$).  Also, all derivatives of the projection operator in
(\ref{Gaaijexpn}) are understood to be evaluated on the degeneracy
manifold at point $(y,0)$.  As for (\ref{GintermsofH}), it can be
expressed in the basis $\ket{Dx;k}$, whereupon it becomes
\begin{equation}
  G^{aa}_{ij}(y,z) = [H^{ab}_i R(\epsilon_0)^2 H^{ba}_j 
  -{\rm h.c.}]+\ldots,
  \end{equation}
and similarly for the other components $(iI)$, $(Ii)$ and $(IJ)$,
where now the derivatives of $H$ are evaluated at $(y,0)$.

To obtain the connection in a neighborhood of $D$ we begin with the
components $F_i$.   These are related to the components $G_{ij}$ of
the curvature by an integral formula like (\ref{FGeqn}), that is,
\begin{equation}
	F^{aa}_i(y,z) = \int_0^1 d\lambda \,
	\lambda z^j\,G_{ji}(y,\lambda z).
	\label{FGijintegral}
	\end{equation}
This is really a special case of (\ref{FGeqn}), which applies to any
parameter space that the Hamiltonian depends on, upon which the
coordinates are $x^\mu$; by interpreting that parameter space as the
transverse space $y={\rm const.}$\ and by replacing $x^\mu$ with
$z^i$, we obtain (\ref{FGijintegral}).  One can also repeat the
derivation of (\ref{FGeqn}) with a changed notation.  This then
implies a version of (\ref{FGexpns}),
\begin{subequations}
\label{FGexpnsij}
\begin{eqnarray}
  F^{aa}_i &=& 0,\\
  \partial_j F^{aa}_i &=& \frac{1}{2} G^{aa}_{ji},\\
  \partial_j \partial_k F^{aa}_i &=&
  \frac{1}{3}[\partial_k G^{aa}_{ji} +
  \partial_j G^{aa}_{ki}],
  \end{eqnarray}
\end{subequations}
etc., where both sides are understood to be evaluated at $(y,0)$.  Now
expressing the curvature in terms of the Hamiltonian, we obtain a
Taylor series for $F_i$ in terms of the coordinates $z^i$,
\begin{equation}
	F^{aa}_i(y,z) = \frac{1}{2}[H^{ab}_j R(\epsilon_0)^2
	H^{ba}_i-{\rm h.c.}]z^j + \ldots,
	\label{Fiexpn}
	\end{equation}
a version of (\ref{FintermsofH}). Here we omit the second order term
(which is available) as it is rather lengthy, and the derivatives on
the right hand side are understood to be evaluated at $(y,0)$.

As for the components $F_I$, we use the integral formula,
\begin{equation}
	F^{aa}_I(y,z) = F^{aa}_I(y,0) + \int_0^1 d\lambda
	\, z^j\,G^{aa}_{jI}(y,\lambda z),
	\label{FGintegralI}
	\end{equation}
which may be compared with (\ref{FGeqn}).  In this formula it is
understood that components of the connection and curvature are taken
with respect to the basis $\ket{Dx;k}$ (while the basis $\ket{px;k}$
was used in (\ref{FGeqn})).  To prove (\ref{FGintegralI}) we express
$G$ in terms of $F$ so that the integral becomes
\begin{eqnarray}
	&&\int_0^1 d\lambda \, z^j\{
	(\partial_j F^{aa}_I)(y,\lambda z) -
	(\partial_I F^{aa}_j)(y,\lambda z)
        \nonumber\\
	&+&[F^{aa}_j(y,\lambda z),F^{aa}_I(y,\lambda z)]\},
\end{eqnarray}
and we use (\ref{Fitransverse}), which causes the commutator and the
second term to vanish.  As for the first term, it is
\begin{equation}
	\int_0^1 d\lambda \,
	\frac{d}{d\lambda}F^{aa}_I(y,\lambda z)
	=F^{aa}_I(y,z) - F^{aa}_I(y,0),
	\end{equation}
which proves (\ref{FGintegralI}).

Now we expand the integrand in (\ref{FGintegralI}) in powers of
$\lambda$ and do the integral to obtain
\begin{equation}
	F^{aa}_I(y,z) = F^{aa}_I(y,0) + z^j\,G^{aa}_{jI}(y,0)
	+\frac{1}{2} z^jz^k\,(\partial_k G^{aa}_{jI})(y,0)
	+\ldots
	\end{equation}
This implies
\begin{subequations}
\label{FGexpnsiI}
\begin{eqnarray}
  \partial_i F^{aa}_I &=& G^{aa}_{iI},\\
  \partial_i \partial_j F^{aa}_I &=&
  \frac{1}{2}[\partial_j G^{aa}_{iI} +
  \partial_i G^{aa}_{jI}],
  \end{eqnarray}
\end{subequations}
where everything is evaluated at $(y,0)$.  Finally, we may express the
connection in terms of the Hamiltonian and its derivatives,
\begin{equation}
  F^{aa}_I(y,z)=F^{aa}_I(y,0) + [H^{ab}_i R(\epsilon_0)^2 H^{ba}_I
  -{\rm h.c.}]z^i + \ldots,
  \label{FIexpn}
  \end{equation}
where again we omit the available second order term due to its length.
Taken with (\ref{Fiexpn}) this provides an expansion of the connection
in a neighborhood of $D$.  We see that the components along $D$ and
those transverse to $D$ can both be expressed in terms of the
curvature, but the expansions are different.  Again, the $bb$-block is
obtained by swapping $a\leftrightarrow b$.  

We make one final comment.   In the case of triatomic molecules in the
electrostatic model the degeneracy manifold $D$ is a one-dimensional
curve inside the section $S$, as noted.   Thus if we use parallel
transport along $D$ to create frame on $D$, the component of $F$ along
$D$ will vanish.   That is, $F^{aa}_I(y,0)=0$, where there is only one
index $I$ since $D$ is one-dimensional.  But then (\ref{Fiexpn}) shows
that the transverse components $F^{aa}_i$ also vanish on $D$.   Thus
we have shown that in the 3-body problem with the electrostatic model
for the electronic Hamiltonian,  there exists a diabatic basis such
that the derivative couplings vanish on the degeneracy manifold.  This
construction only works when the degeneracy manifold is
one-dimensional.

\section{Conclusions}
\label{conclusions}

In this article we have given two versions of a parallel-transported
diabatic basis, one valid in a neighborhood of a point (which is
allowed to lie on a degeneracy manifold or seam), and another which is
valid in a neighborhood of a degeneracy manifold of possibly global
extent.  In both cases we have given Taylor series expansions of the
basis vectors, the derivative couplings and the curvature.  We have
demonstrated the close relationship between the parallel-transported
basis and the singular-value basis, showing that they agree to second
order in a Taylor series expansion about a point.  Our expansion of
the singular-value basis seems to be new.  We have promoted a method
of carrying out these expansions that relies on the projection
operator and that avoids small or vanishing energy denominators or
other singularities.  This approach seems to be new in the literature
on diabatic bases.  We have also exploited integral relationships that
hold between the connection and curvature in the parallel-transported
basis, which are generalizations of Poincar\'e gauge to a non-Abelian
context and which provide a convenient means for computing the
derivative couplings.  Our goal is to give analytic treatments of
connection and curvature in the neighborhood of degeneracy manifolds
that will be useful for future work, including multi-dimensional
Landau-Zener normal forms.

\begin{acknowledgments}
One of us (R.L.) would like to thank Alden Mead for his kind
hospitality and many stimulating and memorable conversations during
the spring of 1993, as well as his support of the present work. The authors would also like to acknowledge many stimulating conversations and much good advice from Enzo Aquilanti, Tucker Carrington, Eric Heller and David Yarkony.  JES was supported by the National Science
Foundation under Grant No. CHE-2102402.
\end{acknowledgments}

\appendix

\section{Some Proofs}
\label{proofs}

Identities involving the projection operators $P(x)$ and $Q(x)$
include $P+Q=1$, $P^2=P$, $Q^2=Q$, $QP=PQ=0$, $P'=P'P+PP'$ and
$PP'P=0$, where prime means differentiation with respect to a
parameter $\lambda$.  

To prove (\ref{Qpsieqn}) we assume (\ref{psitransport}) is true so
that the left hand side of (\ref{Qpsieqn}) can be written
\begin{eqnarray}
  &&\left(\frac{d}{d\lambda}+P'\right)[(1-P)\psi]
  =-P'\psi+(1-P)\psi' +P'(1-P)\psi
  \nonumber\\
  &&\quad =[-P'+(1-P)P' +P'(1-P)]\psi
  \nonumber\\
  &&\quad =(P'-P'P-PP')\psi=0.
  \end{eqnarray}

A similar proof is to show that if $\psi$ satisfies
(\ref{psitransport1}) then $Q\psi$ satisfies (\ref{Qpsieqn}).  We have
\begin{eqnarray}
  &&\left(\frac{d}{d\lambda}+P'\right)[(1-P)\psi] 
  \nonumber\\
  &&=[-P'+(1-P)(P'P-PP')+P'(1-P)]\psi=0.
  \end{eqnarray}
Next we show that if $\psi$ satisfies (\ref{psitransport1}) and
$Q\psi=0$ then $\psi'=P'\psi$.  First, $Q\psi=0$ implies
$\psi=P\psi$.  Next, we are given $\psi'=P'P\psi - PP'\psi$, of which
the first term is $P'\psi$ and the second is $-PP'P\psi=0$.   Thus,
$\psi'=P'\psi$.  

Now we prove (\ref{GammaF}).   We work in any basis for which the
derivatives are defined, which we denote simply by $\ket{x;k}$.  We
are carrying out parallel transport along a curve $x(\lambda)$
according to (\ref{psitransport1}), where the prime means
$d/d\lambda$. 

First we define
\begin{equation}
  F_{\lambda,kl} = \matrixelement{x;k}{\frac{d}{d\lambda}}{x;l}
    =F_{\mu,kl}\,\frac{dx^\mu}{d\lambda}
    =-\left(\frac{d}{d\lambda}\bra{x;k}\right)\ket{x;l}.
    \label{Flambdadef}
    \end{equation}
Next we note that $P\ket{x;k}=\ket{x;k}$ if $k\in A$ and $0$
otherwise.  This implies $\matrixelement{x;k}{P}{x;l}=\delta_{kl}$ if
$k,l\in A$, and $0$ otherwise.  Then by differentiating this we obtain
\begin{equation}
  \matrixelement{x;k}{P'}{x;l}=\begin{cases}
    -F_{\lambda,kl} & \text{if $k\in A$ and $l\notin A$,}\\
    +F_{\lambda,kl} & \text{if $k\notin A$ and $l\in A$,}\\
    0 & \text{otherwise.}
    \end{cases}
    \end{equation}
This then implies
\begin{equation}
  \matrixelement{x;k}{[P',P]}{x;l}=\begin{cases}
    F_{\lambda,kl} & \text{if $k\in A$ and $l\notin A$
      or $k\notin A$ and $l\in A$,}\\
    0 & \text{otherwise.}
    \end{cases}
    \end{equation}

Now we differentiate (\ref{psikexpn}) and substitute into
(\ref{psitransport1}), obtaining
\begin{equation}
  \frac{d\ket{\psi}}{d\lambda}=\sum_{{\rm all}\,k}
  \left[\left(\frac{d}{d\lambda}\ket{x;k}\right)\psi_k
  + \ket{x;k}\frac{d\psi_k}{d\lambda}\right]
  =\sum_{{\rm all}\,k} [P',P]\ket{x;k}\psi_k,
  \end{equation}
or,
\begin{equation}
  \frac{d\psi_k}{d\lambda}=\sum_{{\rm all}\, l}
  \left(\matrixelement{x;k}{[P',P]}{x;l}-F_{\lambda,kl}\right)
  \psi_l.
  \end{equation}
Taking now the cases $k\in A$ and $k\notin A$, we find
\begin{equation}
  \frac{d\psi_k}{d\lambda} =\begin{cases}
    -\sum_{l\in A}F_{\lambda,kl}\,\psi_l & \text{if $k\in A$,}\\
    -\sum_{l\notin A}F_{\lambda,kl}\,\psi_l & 
    \text{if $k\notin A$.}
    \end{cases}
    \end{equation}
Comparing this with (\ref{Gammadef}) we obtain (\ref{GammaF}).

\bibliography{rl.bib}

\end{document}